\documentclass[aps,prd,preprint,superscriptaddress,preprintnumbers]{revtex4-1}

\pdfoutput=1
\usepackage{mathrsfs}
\usepackage{amsfonts}
\usepackage{amsmath,amssymb,bm,bbm}
\usepackage[colorlinks=true,linkcolor=blue,citecolor=green,urlcolor=blue]{hyperref}
\usepackage{epsfig}
\usepackage{graphicx}
\usepackage{slashed}
\usepackage{booktabs}
\usepackage{tabu}
\usepackage[dvipsnames]{xcolor}
\usepackage[normalem]{ulem}
\usepackage{tikz}
\usepackage{soul}
\usetikzlibrary{shapes,arrows}
\usetikzlibrary{positioning}
\usepackage{ulem}

\newcommand{\fref}[1]{Fig.~\ref{f.#1}}

\newcommand{\eref}[1]{Eq.~(\ref{e.#1})}

\newcommand{\tref}[1]{Table~\ref{t.#1}}
\newcommand{\pT}{{p_{\rm T}}}
\newcommand{\pTsq}{{p^2_{\rm T}}}
\newcommand{\pTbf}{{{\bm p}_{\rm T}}}
\newcommand{\diff}[1]{\mathrm{d}#1}
\newcommand{\LN}{\mbox{\tiny \rm LN}}
\newcommand{\W}{{\rm W}}
\newcommand{\FO}{{\rm FO}}
\newcommand{\ASY}{{\rm ASY}}
\newcommand{\Y}{{\rm Y}}
\newcommand{\xb}{x_B}

\begin{document}
\preprint{JLAB-THY-21-3328}

\title{Towards the three-dimensional parton structure of the pion: \\ Integrating transverse momentum data into global QCD analysis}

\author{N. Y. Cao}
\affiliation{Harvard University, Cambridge, Massachusetts 02138, USA}
\author{P. C. Barry}
\affiliation{North Carolina State University, Raleigh, North Carolina 27607, USA}
\affiliation{Jefferson Lab,
	     Newport News, Virginia 23606, USA \\
        \vspace*{0.2cm}
        {\bf Jefferson Lab Angular Momentum (JAM) Collaboration
        \vspace*{0.2cm} }}
\author{N. Sato}
\affiliation{Jefferson Lab,
	     Newport News, Virginia 23606, USA \\
        \vspace*{0.2cm}
        {\bf Jefferson Lab Angular Momentum (JAM) Collaboration
        \vspace*{0.2cm} }}
\author{W. Melnitchouk}
\affiliation{Jefferson Lab,
	     Newport News, Virginia 23606, USA \\
        \vspace*{0.2cm}
        {\bf Jefferson Lab Angular Momentum (JAM) Collaboration
        \vspace*{0.2cm} }}

\begin{abstract}
We perform a new Monte Carlo QCD analysis of pion parton distribution functions, including, for the first time, transverse momentum dependent pion-nucleus Drell-Yan cross sections together with $\pT$-integrated Drell-Yan and leading neutron electroproduction data from HERA.
We assess the sensitivity of the Monte Carlo fits to kinematic cuts, factorization scale, and parametrization choice, and we discuss the impact of the various datasets on the pion's quark and gluon distributions.
This study provides the necessary step towards the simultaneous analysis of collinear and transverse momentum dependent pion distributions and the determination of the pion's \mbox{three-dimensional} structure.
\end{abstract}

\date{\today}
\maketitle

\section{Introduction}
\label{sec.Intro}

The pion is one of the most enigmatic particles in nature, with unresolved questions about its most fundamental properties and behavior.
Unlike all other hadrons, pions play a crucial role in QCD as the nearly massless Goldstone bosons associated with the breaking of chiral symmetry.
At the same time, as strongly interacting bound states of quarks, antiquarks, and gluons (or partons), they reveal a familiar spectrum of momentum distributions at high energies, as do other hadrons and nuclei.
Understanding how these contrasting manifestations of the same $q\bar q$ bound state arise dynamically at different energy scales from first principles remains a major challenge in QCD, and progress towards the ultimate resolution of the partonic-hadronic duality of the pion at present requires phenomenological input~\cite{Melnitchouk:2005zr, Duality03}.

Recently the Jefferson Lab Angular Momentum (JAM) collaboration performed the first Monte Carlo global QCD analysis of pion parton distribution functions (PDFs)~\cite{Barry18} constrained by the traditional Drell-Yan dilepton production data in pion-nucleus scattering and data on leading neutron production in inclusive lepton-proton collisions.
While the Drell-Yan data from CERN~\cite{NA10} and Fermilab~\cite{E615} provide constraints on the valence quark PDFs at intermediate and large values of the parton momentum fraction $x$ in the pion, the leading neutron data~\cite{H1, ZEUS} were important in establishing the role played by gluons in the pion at low~$x$. 
Global QCD studies of Drell-Yan and prompt photon data were previously performed in Refs.~\cite{Owens84, ABFKW89, SMRS, GRV, GRS, Wijesooriya05, xFitter20} in the context of single-fit analyses.
In addition to the phenomenological studies, there has also been considerable interest recently in computing the pion's valence quark distribution in lattice QCD~\cite{Sufian:2019bol, Izubuchi:2019lyk, Joo:2019bzr, Sufian:2020vzb, Karthik:2021qwz}.

From another direction, recent progress in the theory of transverse momentum dependent (TMD) distributions has inspired exploratory studies of the transverse momentum structure of partons in the pion~\cite{Vladimirov:2019bfa}, which have attempted to describe the $\pT$ dependence of the differential pion-induced Drell-Yan cross section.
The theoretical formalism developed in the 1980s by Collins, Soper and Sterman (CSS)~\cite{Collins:1984kg} allows one to describe $\pT$-dependent observables within the so-called ``$\W$+$\Y$'' framework (where $\pT$ is the transverse momentum of the virtual photon relative to the beam axis of the colliding hadrons).
In this construction, the ``$\W$'' term is computed in terms of TMD parton distributions in the region where $\pT \ll Q$, where $Q$ is the invariant mass of the lepton pair.
At such kinematics, the transverse momentum of the virtual photon is sensitive to the intrinsic transverse momentum of partons in the hadron.
Crucially, the TMDs here are sensitive to the vacuum expectation value of Wilson lines, which is expected to be independent of the type of hadron involved.
Confirmation of such universality for all hadrons would be one of the most important validations of QCD factorization theorems and their predictive power.

On the other hand, in regions of large transverse momenta $\pT \sim Q$, the spectrum is dominated by hard QCD radiation in which the photon recoils from hard radiated partons.
In such situations, the cross sections can be described by collinear factorization involving collinear PDFs.
In practice, the CSS framework has been very successful in applications in collider environments such as $W$ and $Z$ production in large-$\pT$ $p\bar p$ and $pp$ collisions at the Tevatron and LHC, respectively.
One of the main reasons for this success is that collider setups offer sufficient phase space for the final states to allow perturbative QCD calculations to provide good descriptions of the data, in contrast to low-energy reactions, such as those in fixed-target experiments, where the applicability of the CSS framework is more limited.
In Ref.~\cite{Bacchetta:2019tcu}, it was found, for example, that even with the inclusion of ${\cal O}(\alpha_s^2)$ corrections, the theoretical predictions in the large-$\pT$ region significantly underestimate the measured cross section for low-energy experiments.
Similar trends have been observed in semi-inclusive deep-inelastic scattering~\cite{Gonzalez-Hernandez:2018ipj, Wang:2019bvb, Boglione:2019nwk}.

The situation is somewhat different in the small-$\pT$ region, where several analyses have attempted to describe the data in terms of TMDs~\cite{Bacchetta:2019sam, Delcarro:2018lbr, Scimemi:2019cmh}.
These efforts have been partially successful, in that the analyses have been able to describe hadron multiplicities at small transverse momenta, making cuts on data with $\pT$ above some maximum value.
A globally successful analysis, on the other hand, would seek to include the entire $\pT$ spectrum, describing the data in terms of the full $\W+\Y$ treatment.

Such an analysis would be important for several reasons, including the fact that, on formal grounds, the applicability of the $\W$-term is limited for regions where $\pT/Q \ll 1$. Despite that, data beyond the region of applicability, up to $\pT/Q \sim 1$, are still often included in low-energy TMD analyses, thereby potentially biasing the extraction of the TMDs.
Additionally, the large transverse momenta give access to PDFs at large parton momentum fractions, making it more feasible, for instance, to probe gluon distributions at large $x$, through gluon initiated subchannels that arise at the same order in $\alpha_s$ as quark channels.

Given that TMDs are now considered to be a key tool in describing the structure of hadrons, the question arises of whether it can be confirmed that collinear factorization can serve as an equally reliable tool for describing inclusive particle production at high $\pT$.
Motivated by the state of the art of TMD phenomenology, as well as the broad interest in exploring the inner structure of pions and the recent success in extracting pion TMDs, in this paper we analyze the large transverse momentum region of Drell-Yan lepton-pair production in fixed target pion-nucleus scattering.
We will demonstrate that, contrary to proton-induced Drell-Yan reactions, in the case of the pion, it is indeed possible to describe the large transverse momentum region and include the available $\pT$-dependent data consistently within a global analysis of pion PDFs constrained by $\pT$-integrated data, as in Ref.~\cite{Barry18}.
This analysis is a first step towards a full combined analysis of all available pion-induced data at low and high $\pT$ to simultaneously determine collinear and TMD pion PDFs within the same theoretical framework.

In Sec.~\ref{sec.Theory} of this paper, we outline the theoretical formalism used in our analysis to describe $\pT$-dependent and $\pT$-integrated Drell-Yan lepton-pair production cross sections in pion-nucleus scattering, along with the inclusive electroproduction of forward neutrons in electron-proton collisions.
We perform the analysis within the JAM Monte Carlo fitting framework, used in previous successful analyses of pion PDFs~\cite{Barry18}, unpolarized nucleon PDFs and fragmentation functions~\cite{JAM19, Sato:2016wqj}, and spin-dependent nucleon PDFs~\cite{Ethier:2017zbq}, as well as the nucleon's transversity distributions~\cite{Lin:2017stx} and other TMD functions~\cite{Cammarota:2020qcw}.
The datasets used in this analysis are described in Sec.~\ref{s.Datasets}, which include Drell-Yan lepton pair production cross sections in pion-nucleus scattering from CERN~\cite{NA10} and Fermilab~\cite{E615}, including both $\pT$-differential and $\pT$-integrated data, as well as leading neutron electroproduction cross sections from HERA~\cite{H1, ZEUS}.
In Sec.~\ref{s.methodology} we describe the Monte Carlo methodology used to extract the PDFs through Bayesian inference.
The results for the pion's valence and sea quark and gluon PDFs are presented in Sec.~\ref{s.results}, where we examine the dependence of the analysis on kinematic cuts, factorization scale, and choice of PDF parametrization.
In Sec.~\ref{s.PionVsProton} we compare the results of the analysis with PDFs in the proton, and finally we summarize our findings and discuss future extensions of this work in Sec.~\ref{s.Outlook}.

\section{Theoretical framework}
\label{sec.Theory}

In this section we present the theoretical framework used in the present analysis of PDFs in the pion, including the relevant formulas for $\pT$-differential and $\pT$-integrated cross sections for the Drell-Yan process in pion-nucleus scattering~\cite{DY}, along with the semi-inclusive forward neutron production cross sections in electron-proton collisions.
We provide only the relevant formulas needed for our immediate calculations; more details about the formalism for both types of reactions can be found in Refs.~\cite{Owens, McKenney, Barry18, PionResum}.

\subsection{Drell-Yan lepton-pair production}
\label{ssec.dy}

Within the CSS framework the $\pT$-differential cross section for the inclusive production of a lepton pair $\ell^+ \ell^-$ in the high-energy collision of hadrons $A$ and $B$, $A\, B \to \ell^+ \ell^- X$, can be expressed schematically in the form
\begin{align}
\frac{\diff^3 \sigma^{\mbox{\tiny \rm DY}}}{\diff Q^2\, \diff y\, \diff p^2_{\rm T}}
&= \W+\FO-\ASY+{\cal O}\bigg(\frac{m^2}{Q^2}\bigg),
\label{e.w+y}
\end{align}
where ``W'', ``FO'' and ``ASY'' refer to the W-term, the fixed-order term, and the asymptotic contribution, respectively~\cite{Collins:2011zzd}, and the kinematic variables are the lepton pair's invariant mass squared $Q^2$, rapidity $y$, and transverse momentum $\pT$.
The corrections to the formula~(\ref{e.w+y}) appear in the form of ratios of masses $m$ to the large scale $Q$.

The $\W$-term in Eq.~(\ref{e.w+y}) is computable in terms of TMD PDFs, and is a valid approximation to the cross section in the low-$\pT$ region, $\pT/Q\ll 1$, with errors ${\cal O}(\pT/Q)$.
In contrast, the $\FO$ term is applicable in the region where $\pT\sim Q$, with corresponding errors ${\cal O}(m/\pT)$.
Finally, the asymptotic term $\ASY$ in \eref{w+y} is, by construction, the large transverse momentum approximation to the $\W$-term, as well as the small transverse momentum approximation to the $\FO$-term. 
Defining the ``$\Y$-term'' by $\Y \equiv \FO-\ASY$, in the small-$\pT$ limit one has
    $\lim_{\pT/Q\to 0} \Y=0$
and the cross section is dominated by the $\W$-term, while at large $\pT$ by construction
    $\lim_{\pT\sim Q} (\W-\ASY)=0$
and the cross section is dominated by the $\FO$-term.
In the present analysis, we will focus on the latter region and examine in detail the $\FO$-term.
Further discussion about the $\W$ and $\ASY$ contributions can be found at Ref.~\cite{Collins:1984kg}.

\begin{figure}
    \centering
    \includegraphics[width=0.35\textwidth]{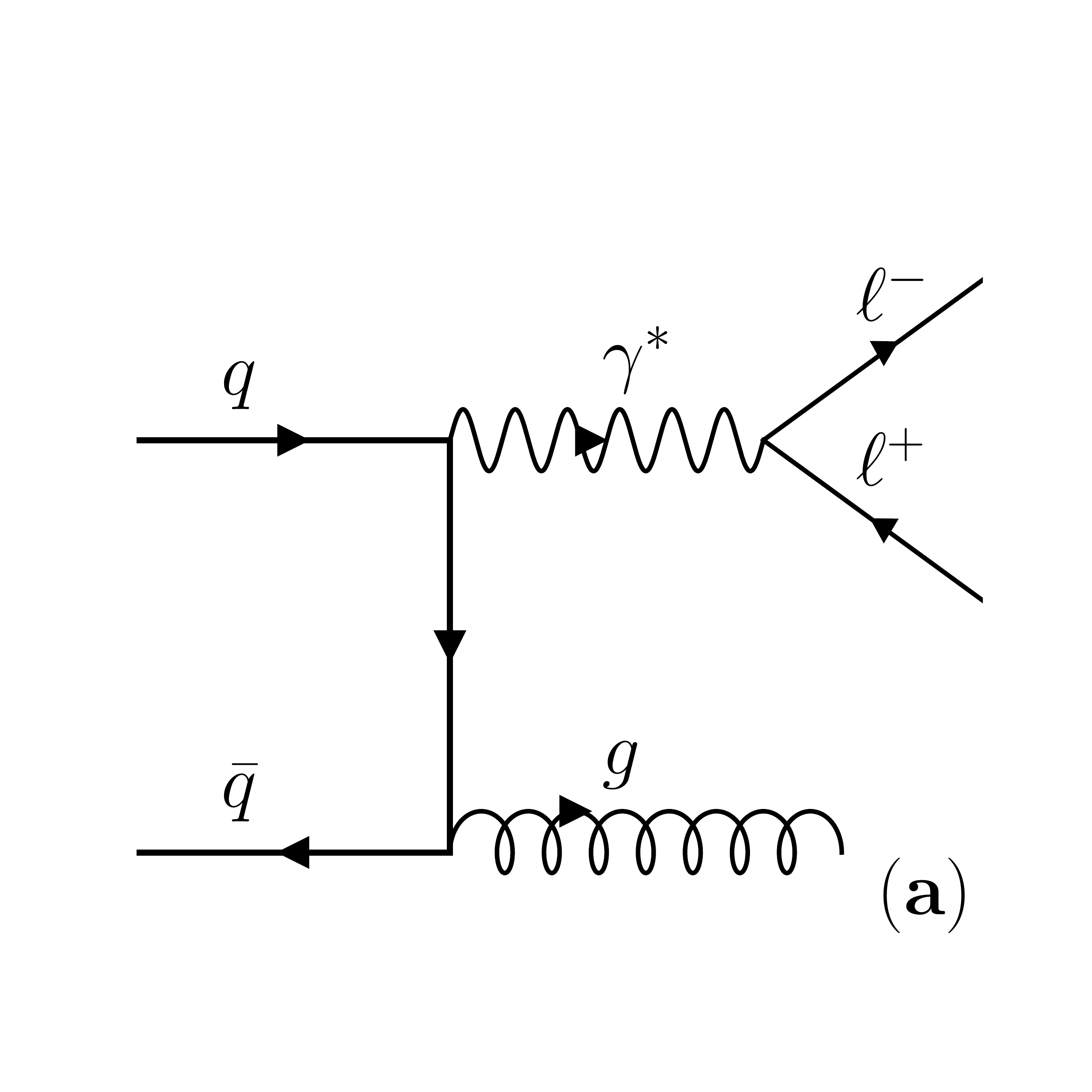}\qquad\quad
    \includegraphics[width=0.35\textwidth]{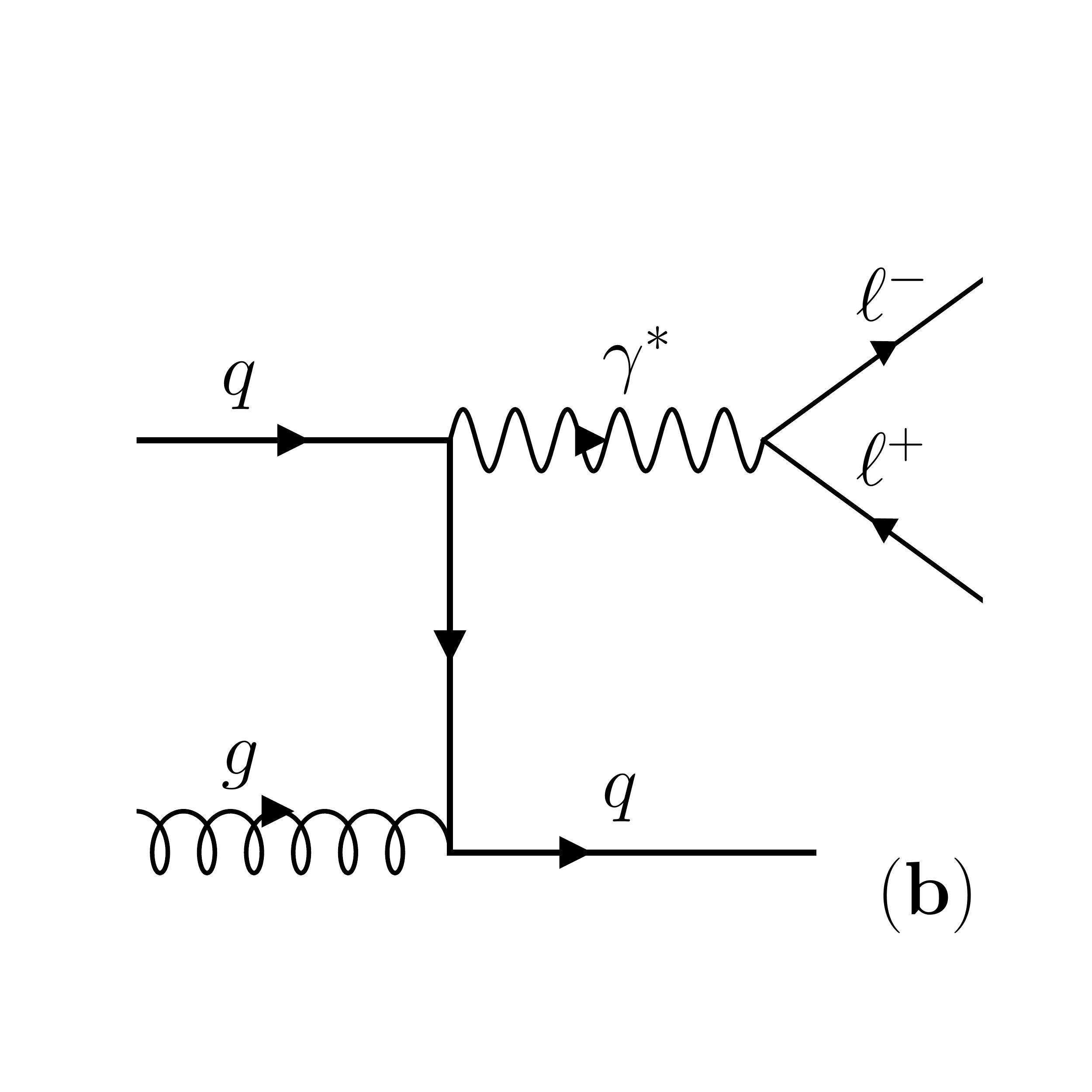}
    \caption{Examples of LO diagrams for the large transverse momentum region in Drell-Yan lepton-pair production for the $q\bar{q}$ channel {\bf (a)} and $qg$ channel {\bf (b)}.}
    \label{f.DYpTLO}
\end{figure}

To produce a virtual photon with a large transverse momentum, the photon must recoil against a hard radiation, examples of which are shown in Fig.~\ref{f.DYpTLO}.
Since the scale of such radiation is large, it cannot be associated with an intrinsic transverse momentum of quarks and gluons in the nucleon, but instead involves hard radiation, which is calculable perturbatively in QCD.
It will be convenient to parametrize the virtual photon's four-momentum $q$ in terms of its invariant mass $Q$, transverse momentum $\pTbf$ and rapidity $y$,
\begin{eqnarray}
q^\mu
&=& \left(  Q_{\rm T} \cosh y;\, \pTbf,\, Q_{\rm T} \sinh y \right),
\end{eqnarray}
where $Q_{\rm T} \equiv \sqrt{Q^2+\pTsq}$ is the transverse mass of the photon.
Schematically, the $\FO$-term can then be written, to all orders in the strong coupling $\alpha_s$, as
\begin{eqnarray}
\FO(y,Q^2,\pT)
&=&
\sum_{a,b} \int \diff x_a\, \diff x_b\,
H_{a,b}^{\mbox{\tiny \rm FO}}(x_a,x_b,y,p_{\rm T},Q^2,\mu^2)\,
f_a^A(x_a,\mu^2)\,
f_b^B(x_b,\mu^2),
\label{e.FO}
\end{eqnarray}
where $f_a^A$ ($f_b^B$) is the collinear PDF for flavor $a$ ($b$) in hadron $A$ ($B$), evaluated at parton momentum fraction $x_a$ ($x_b$) at a renormalization scale $\mu$, and $H_{a,b}^{\mbox{\tiny \rm FO}}$ is the perturbatively calculable short-distance cross section.
Note that the integration limits in Eq.~(\ref{e.FO}) depend on all three variables $y$, $Q^2$ and $\pT$.

The production of a large transverse momentum photon can proceed via two types of partonic hard scattering subprocesses,
    $q\bar{q} \to \ell^+ \ell^- g$ (Fig.~\ref{f.DYpTLO}(a)) and $qg \to \ell^+ \ell^- q$ (Fig.~\ref{f.DYpTLO}(b)),
which start at ${\cal O}(\alpha_s)$ in the QCD coupling constant.
At this order, the final state phase space constrains one of the parton momentum fraction integrals so that $x_a$ and $x_b$ are not independent, but are related~by
\begin{align}
x_b = \frac{x_a\, z_- - \tau}{x_a - z_+},
\end{align}
where 
\begin{align}
z_\pm = \sqrt{\Big( \tau + \frac{\pTsq}{s}\Big)}\ e^{\pm y}\, , 
\hspace{0.7cm}
\tau = \frac{Q^2}{s},
\end{align}
and the minimum value of $x_a$ is given by
\begin{align}
x_a^{\rm min} = \frac{z_+-\tau}{1-z_-}.
\end{align}
As $\pT$ increases, the lower limit of $x_a^{\rm min}$ increases, so that the cross section becomes increasingly more sensitive to PDFs with large parton momentum fractions.
While the perturbative QCD calculations for $H^{\mbox{\tiny \rm FO}}_{a,b}$ are known up to ${\cal O}(\alpha_s^2)$~\cite{Arnold89}, in the present analysis we will restrict ourselves to ${\cal O}(\alpha_s)$ in order to test the applicability of the large transverse momentum region using the lowest order in pion induced Drell-Yan reactions.

It is instructive to contrast the theoretical calculation for the $\pT$-differential cross section~(\ref{e.w+y}) with the corresponding cross section integrated over all $\pT$.
In this case the $\pT$-integrated cross section can be written in the more familiar form of collinear factorization~as
\begin{align}
\frac{\diff^2 \sigma^{\mbox{\tiny \rm DY}}}{\diff Q^2 \diff y}
=& \sum_{a,b}
     \int
     \diff x_a\,
     \diff x_b\,
     H_{a,b}^{\mbox{\tiny \rm DY}}(x_a,x_b,y,Q^2,\mu^2)\,
     f_a^A(x_a,\mu^2)\,
     f_b^B(x_b,\mu^2),
\label{e.dyint}
\end{align}
where in this case the limits of integration no longer depend on $\pT$.
The hard cross section $H_{a,b}^{\mbox{\tiny \rm DY}}$ starts at ${\cal O}(\alpha_s^0)$, and in our analysis we compute corrections up to ${\cal O}(\alpha_s)$.
Our study is the first attempt to include both $\pT$-differential and $\pT$-integrated pion-nucleus Drell-Yan data~\cite{NA10, E615} on the same footing, taking advantage of the fact that the $\pT$-dependent cross sections provide access to a larger region of parton momentum fractions relative to the $\pT$-integrated~case.

\subsection{Leading neutron electroproduction}
\label{ssec.ln}

In addition to the inclusive lepton-pair production cross sections in pion-nucleus scattering in Eqs.~(\ref{e.w+y}) and (\ref{e.dyint}), following Ref.~\cite{Barry18}, we supplement the analysis with data from leading neutron~(LN) electroproduction in $ep$ collisions at HERA~\cite{H1, ZEUS} in the very forward region, $e p \to e n X$, in order to better constrain the pion PDFs at small values of parton momentum fraction $x$ in the pion.
By detecting a forward neutron and lepton in the final state, the hard scattering is expected to occur between the exchanged virtual photon (four-momentum~$q = \ell - \ell'$, where $\ell$ and $\ell'$ are the incident and scattering lepton momenta) and the pion in the limit where the momentum transfer squared $t = (p-p')^2$ from the proton~($p$) to the final neutron~($p'$) is small~\cite{Sullivan, Thomas:1983fh}.

The triply differential LN cross section is typically parametrized by the LN structure function, $F_2^{\LN}$, which is a function of the Bjorken scaling variable $\xb = Q^2/2 p \cdot q$, the invariant mass squared of the virtual photon, $Q^2 = -q^2$, and the longitudinal momentum fraction carried by the detected neutron, $x_L = p' \cdot q/p \cdot q$,
\begin{equation}
\frac{\diff^3 \sigma^{\LN}}{\diff \xb\, \diff Q^2\, \diff x_L}
= \frac{4\pi\alpha^2}{\xb\, Q^4}
  \Big( 1 - y_e + \frac{y_e^2}{2} \Big)\,
  F_2^{\LN}(\xb,Q^2,x_L),
\label{eq.LNxsec}
\end{equation}
where $y_e = q \cdot p / \ell \cdot p \approx Q^2/x s$ is the lepton inelasticity, and $s=(\ell+p)^2$ is the total center of mass energy ($\sqrt{s}$ $\approx 300$~GeV at HERA kinematics).

In the limit where $|t| \to 0$ and $\bar{x}_L \equiv 1-x_L \to 0$, one expects the charge exchange process $\gamma^* p \to n X$ to be dominated by the emission and absorption of pions~\cite{Sullivan, Thomas:1983fh, Melnitchouk:1992yd, DAlensio00, Kopeliovich12}.
In this region, the chiral symmetry properties of QCD suggest that the peripheral structure of the nucleon can be described through interactions between the probe and the ``cloud'' of pseudoscalar mesons associated with the long-range structure of the nucleon.
Formally, the effects of this structure on the total LN cross section can be computed within chiral effective field theory, by matching twist-two partonic operators with corresponding hadronic operators having the same quantum numbers~\cite{ChenPRL, Arndt:2001ye, Moiseeva13}.

This leads to a representation for the matrix elements of the twist-two operators between nucleon states in terms of matrix elements of the hadronic operators and matrix elements of the twist-two operators between the hadronic states, which can in turn be expressed through moments of the corresponding light-cone correlation functions.
The validity of this construction for all moments implies its veracity also in momentum space, which allows the LN structure function to be written as
\begin{equation}
F_2^{\LN}(\xb,Q^2,x_L)\,
=\, 2 f_{\pi N}(\bar{x}_L)\, F_2^\pi (x_\pi,Q^2)\,
+\, \textrm{multi-pion~corrections},
\label{eq.F2LN3}
\end{equation}
where $f_{\pi N}$ is the light-cone momentum distribution of pions in the nucleon (or nucleon $\to$ nucleon $+$ pion splitting function)~\cite{Burkardt13, Salamu15, XWangPRD, Salamu:2018cny, Salamu:2019dok}, and $F_2^\pi$ is the inclusive structure function of the pion evaluated at $x_\pi = \xb / \bar{x}_L$.
Note the factor $2$ in Eq.~(\ref{eq.F2LN3}) is an isospin factor for the application of the $p\to n \pi^+$ fluctuation, whose distribution is related to the $p\to p \pi^0$ fluctuation by $f_{\pi^+ n} =2f_{\pi^0 p}=2f_{\pi N}$.
Within the chiral effective theory framework, at the one-pion loop level the splitting function $f_{\pi N}$ is given by~\cite{Sullivan, Thomas:1983fh, Burkardt13, Salamu15, XWangPRD, Salamu:2018cny, Salamu:2019dok, Holtmann:1996be}
\begin{equation}
f_{\pi N}(\bar{x}_L)
= \frac{g_A^2 M^2}{(4\pi f_\pi)^2} \int \diff k_\perp^2
  \frac{\bar{x}_L \big[ k_\perp^2+\bar{x}_L^2 M^2 \big]}{(1-\bar{x}_L)^2 D_{\pi N}^2}\,
  \mathcal{F}^2(\bar{x}_L,k_\perp^2),
\label{eq.splfunc}
\end{equation}
where $D_{\pi N}$ is the pion propagator,
\begin{equation}
D_{\pi N} \equiv t - m_\pi^2
= -\frac{1}{1-\bar{x}_L} \big[ k_\perp^2 + \bar{x}_L^2 M^2 + (1-\bar{x}_L) m_\pi^2 \big],
\label{eq.DpiN}
\end{equation}
and $g_A=1.267$ is the axial charge, $f_\pi=93$~MeV is the pion decay constant,
and $M$ and $m_\pi$ are the nucleon and pion masses, respectively.
The function $\mathcal{F}$ regulates the ultraviolet divergences as $k_\perp^2 \to \infty$, and it is determined phenomenologically in terms of a cutoff mass parameter, $\Lambda$~\cite{McKenney, Barry18, Z1}.
The pion structure function $F_2^\pi$ can be written in the standard collinear factorized form,
\begin{eqnarray}
F_2^\pi(x_\pi,Q^2)
&=& \sum_i \int_{x_\pi}^1 \diff \xi\,
H_i^{{\mbox{\tiny \rm DIS}}}(\xi,\mu^2,Q^2)\,
f_i^\pi(x_\pi/\xi,\mu^2),
\end{eqnarray}
in terms of the PDFs of flavor $i$ in the pion, $f_i^\pi$, and the inclusive DIS hard scattering function, $H_i^{{\mbox{\tiny \rm DIS}}}$.

Having summarized the theoretical formulas relevant for our global PDF study, in the next sections we focus on the details of the datasets that will be fitted, and the methodology employed for the Monte Carlo analysis.

\section{Datasets}
\label{s.Datasets}

\begin{figure}[t]
\includegraphics[width = 0.8\textwidth]{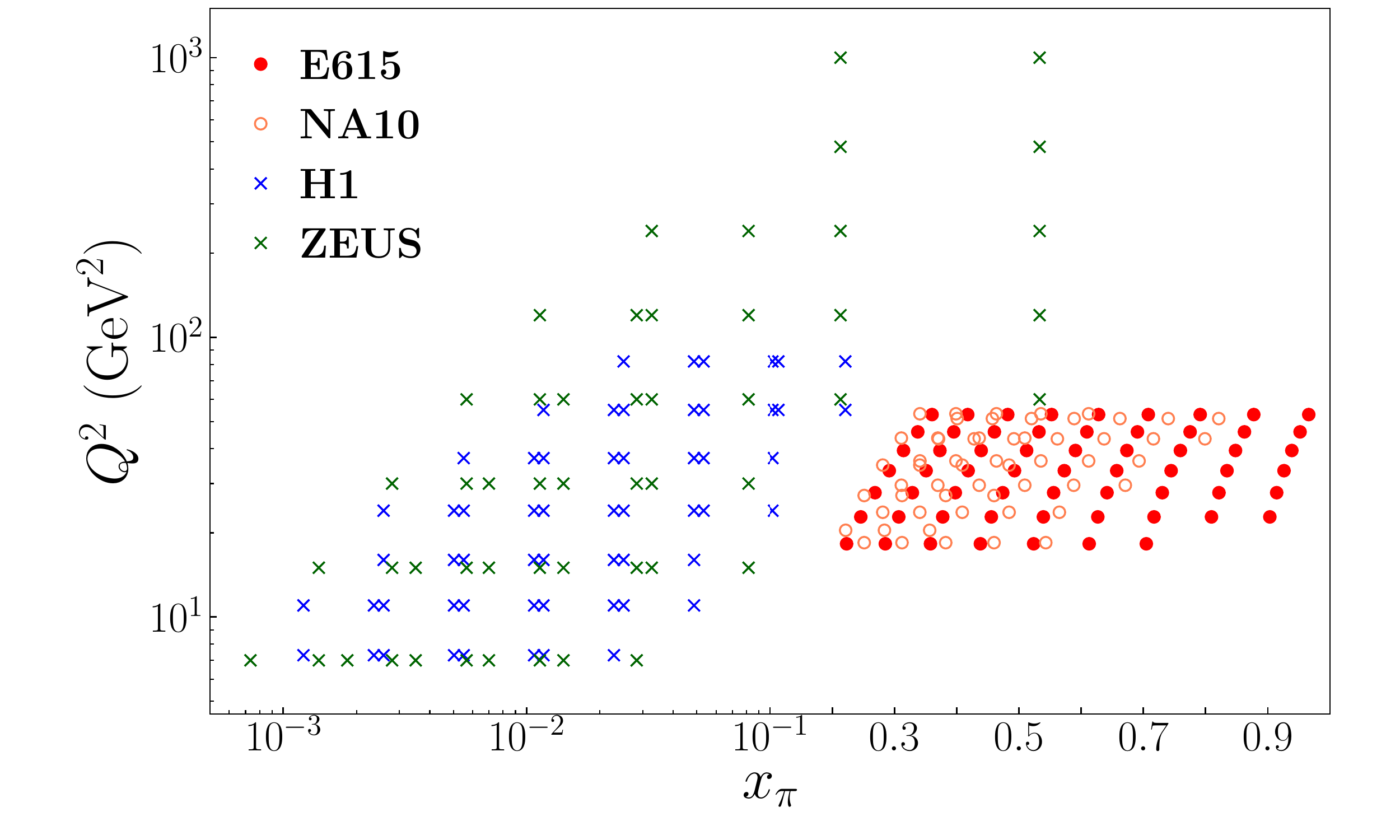}
\vspace*{-0.5cm}
\caption{Kinematic coverage in $Q^2$ and $x_\pi$ of the NA10~\cite{NA10} and E615~\cite{E615} Drell-Yan datasets, where $x_\pi = \frac12 (x_F + \sqrt{x_F^2+4\tau})$, and the H1~\cite{H1} and ZEUS~\cite{ZEUS} leading neutron electroproduction data, for which $x_\pi = \xb/\bar{x}_L$.  The NA10 datasets include two different pion beam energies, 194 and 286~GeV (see Table~\ref{t.chi2}).}
\label{f.kin}
\end{figure}

Following the previous JAM pion PDF analysis~\cite{Barry18}, we fit pion-tungsten Drell-Yan lepton-pair production cross sections from the NA10 experiment at CERN~\cite{NA10} and the E615 experiment at Fermilab~\cite{E615}, together with LN electroproduction data from the ZEUS~\cite{ZEUS} and H1~\cite{H1} experiments at the HERA $ep$ collider.
In addition, we include, for the first time, data on the $\pT$-differential Drell-Yan cross sections from E615, which have not been utilized in global PDF analyses to date.
The kinematic coverage for the Drell-Yan and LN datasets is shown in Fig.~\ref{f.kin}.

The Drell-Yan cross sections are typically presented in terms of $\sqrt{\tau}=\sqrt{Q^2/s}$, where $Q^2$ is the invariant mass squared of the virtual photon and $s$ is the total incoming center of mass energy, and the Feynman scaling variable
    $x_F = x_\pi - x_N =  2\sqrt{\tau}\sinh{y}$,
where $x_\pi = \sqrt{\tau} e^{y} = \frac12 (x_F + \sqrt{x_F^2+4\tau})$
and $x_N = \sqrt{\tau} e^{-y}$ are the nuclear scaling variable per nucleon.
At leading order (LO) in $\alpha_s$, $x_\pi$ and $x_N$ coincide with the parton momentum fractions in the pion beam and nucleon target, respectively.
Both the NA10 and E615 experiments utilized a $\pi^-$ beam incident on a tungsten nucleus.
Pion beam energies of 194 and 286~GeV were used in the NA10 experiment, while the E615 experiment used a $\pi^-$ beam energy of 252~GeV.
The extraction of the pion PDFs from the Drell-Yan data naturally requires knowledge of the PDFs of the target tungsten nuclei.
In our analysis we use central values for the tungsten PDFs from the nuclear PDF analysis by Eskola {\it et~al.}~\cite{EPPS16}.
Using other nuclear PDF sets, such as from the nCTEQ group~\cite{nCTEQ15}, leads to indistinguishable results on pion PDFs or goodness of fit.

To ensure that the Drell-Yan lepton-pair production process is dominated by leading-twist partonic subprocesses, we avoid contributions from the $J/\psi$ and $\Upsilon$ resonances by only selecting events with $m_{J/\psi} \lesssim Q \lesssim m_\Upsilon$.
Furthermore, since with increasing $Q$ the available phase space for the hard factorization becomes smaller, to avoid edges of phase space we restrict the kinematical reach at a fixed $s$ to $Q$ values that are not too large.
In practice, we find these conditions to be optimally satisfied for the $Q$ range given by $4.16 < Q < 7.68$~GeV.
The upper limit is slightly more conservative than that which was adopted in Ref.~\cite{Barry18} in order to avoid increasing deviations between data and theory in the highest-$Q$ bin.

For the range of $x_F$ considered, to avoid contamination from exclusive channels and remain within the region where factorization is valid, we implement the same restrictions as in Ref.~\cite{Barry18}, namely, $x_F < 0.9$.
To test if a smaller $x_F$ would be more appropriate, we also performed fits to the Drell-Yan data with $x_F^{\rm max}$ cuts of 0.6, 0.7, and 0.8, but we found that the $\chi^2$ and resulting PDFs were essentially unchanged from the results with $x_F^{\rm max} = 0.9$.
To retain the largest number of data points that can be simultaneously described in the global analysis, we present results for data in the range $0 < x_F < 0.9$.
This leaves 61 data points from the E615 experiment and 56 from NA10, for a total of 117 data points representing the $\pT$-integrated Drell-Yan cross section.

The $\pT$-dependent lepton-pair production cross sections are from the same E615 Drell-Yan pion-tungsten experiment \cite{E615}, which measured the $\pT$ spectra both integrated over $x_F$,
    $\diff^2\sigma^{\mbox{\tiny \rm DY}}/\diff Q \diff \pT$,
and integrated over $Q^2$,
    $\diff^2\sigma^{\mbox{\tiny \rm DY}}/\diff x_F \diff \pT$.
The $\pT$-dependent E615 data extend up to $p^{\rm max}_{\rm T} = 4.875$~GeV, which is $\approx 25\%$ of the maximum $\pT$ allowed for the $Q^2$ ranges used, and the data are still far from the kinematic edge.
The lower limit on $\pT$ is taken to ensure the applicability of the collinear approximation for the FO term in Eq.~(\ref{e.FO}), and the value
    $p^{\rm min}_{\rm T} = 2.7$~GeV
was determined phenomenologically, as discussed in Sec.~\ref{ssec.quality} below.
For the $x_F$-integrated cross section, the fitted data span the range
    $4.3 < Q <7.2$~GeV,
with a total of 34 data points, while for the $Q$-integrated cross section (which is integrated over $4.05 < Q < 8.55$~GeV) data for $x_F \lesssim 0.9$ are included, with a total of 49 data points.

For the LN electroproduction data, the H1 experiment at HERA~\cite{H1} measured the $F_2^{\LN}$ structure function, related to the triply differential LN cross section in Eq.~(\ref{eq.LNxsec}), while the ZEUS Collaboration~\cite{ZEUS} measured the ratio of LN to inclusive cross sections,
\begin{equation}
r(\xb,Q^2,x_L)
= \frac{\diff^3\sigma^{\mbox{\tiny LN}}/\diff \xb\, \diff Q^2 \diff x_L}
       {\diff^2\sigma^{\rm   }/\diff \xb\, \diff Q^2}\,
  \Delta x_L,
\label{e.ZEUSrxsec}
\end{equation}
for $x_L$ bin sizes $\Delta x_L$.
For the denominator of Eq.~(\ref{e.ZEUSrxsec}) we use the JAM19 PDFs~\cite{JAM19} to compute the inclusive proton cross sections.
As seen in Fig.~\ref{f.kin}, the LN data extend to much lower values of $x_\pi$ than the Drell-Yan data, $x_\pi \gtrsim 10^{-3}$, allowing more direct determination of the pion's sea quark and gluon distributions.
Indirect constraints on the gluon PDFs are obtained from $Q^2$ evolution over the large range of $Q^2$ values covered by the HERA data, from $Q^2 \sim$~a few GeV$^2$ to $10^3$~GeV$^2$.
For the $x_L$ dependence, we restrict the analysis to the region of large $x_L$, where most of the incident proton's longitudinal momentum is carried away by the detected neutron, and one can approximate the LN cross section by the exchange of a soft pion~\cite{DAlensio00, Kopeliovich12}.
We follow the systematic studies of $\chi^2$ performed as a function of the $x_L$ in Refs.~\cite{McKenney, Barry18} and apply the cut $x_L > 0.8$ found there.

\section{Monte Carlo Methodology}
\label{s.methodology}

In this section we describe the analysis procedure used 
to extract the pion PDFs through Bayesian inference.
The basic strategy follows previous JAM analyses of pion~\cite{Barry18} and proton~\cite{JAM19, Ethier:2017zbq} PDFs, with some improvements specific to the present application.

\subsection{Parameter inference}
\label{s.parametrization}

As in earlier global PDF analyses, we use the following standard template for the functional form of the nonperturbative distributions, which are parametrized at the scale $\mu_0^2$,
\begin{equation}
T_i(x,\mu_0^2)
= \frac{N_i\, x^{a_i} (1-x)^{b_i} (1 + c_i \sqrt{x} + d_i x)}
       {B[a_i+2,b_i+1] + c_i B[a_i+\tfrac52,b_i+1] + d_i B[a_i+3,b_i+1]},
\label{e.PDF}
\end{equation}
where $B$ is the Euler beta function, and the normalization $N_i$ is defined to correspond to the average momentum fraction of the pion carried by parton $i$. 
The input scale is defined as $\mu_0=m_c=1.27~{\rm GeV}$.
For the pion PDFs we parametrize 3 degrees of freedom: the valence quark distribution, $q_v^\pi$, the sea quark distribution, $q_s^\pi$, and the gluon distribution, $g^\pi$.
All of the light-quark PDFs can be expressed in terms of the distributions $q_v^\pi$ and $q_s^\pi$, using isospin and SU(3) flavor symmetry. 
In particular, for the valence quark distribution we have
    $q_v^\pi \equiv u_v^{\pi^+} 
    = u^{\pi^+} - \bar{u}^{\pi^+} = \bar{d}_v^{\pi^+}
    = \bar{u}_v^{\pi^-} = d_v^{\pi^-}$,
while for the sea quarks we take
    $q_s^\pi \equiv \bar{u}^{\pi^+} = d^{\pi^+} = s^{\pi^+} = \bar{s}^{\pi^+}$.
The heavy quark flavors in our analysis are generated perturbatively via QCD evolution by solving the DGLAP equations using the zero-mass variable flavor number scheme evolved up to next-to-leading-logarithmic accuracy.

In our numerical analysis we have found that fitting the available data with the 5-parameter form in \eref{PDF} gives essentially no improvement and goodness of fit compared with fitting with 3 parameters per flavor, with $c_i$ and $d_i$ set to zero.
Furthermore, we have explored parametrizing the PDFs through the addition of two template shapes in order to increase flexibility.
Such fits tend to give gluon PDFs with a negative valley at intermediate $x$ values without a significant improvement in $\chi^2$, which suggests the possibility of overfitting.
To keep the number of free parameters to a minimum, we choose to use the form $T_i$ in \eref{PDF} with $c_i = d_i=0$.
With this setup, 9 shape parameters are used to describe the valence, sea and gluon PDFs, and with additional constraints from the valence quark number rule, which fixes $N_v$, and the momentum sum rule, which fixes $N_s$, we are left with a total of 7 shape parameters to fit.

The free parameters are inferred by sampling the Bayesian posterior distribution 
\begin{equation}
\mathcal{P}(\bm{a}|{\rm data}) 
\sim \mathcal{L}({\rm data}|\bm{a})\, \pi(\bm{a}),
\label{e.Bayes}
\end{equation}
where $\bm{a} = \{ N_i, a_i, b_i, \ldots \}$ represents the vector of parameters, and the likelihood function ${\cal L}$ is given by
\begin{equation}
\mathcal{L}({\rm data}|\bm{a}) 
= \exp{\Big(-\frac{1}{2} \chi^2(\bm{a},{\rm data})\Big)}.
\label{e.likelihood}
\end{equation}
The argument of the exponential in (\ref{e.likelihood}) is given by the $\chi^2$ function,
\begin{align}
\chi^2(\bm{a},{\rm data}) 
&= \sum_e \left( \sum_i 
\bigg[
\frac{d_i^{\, e} - \sum_k r^e_k\, \beta^{\, e}_{k,i} - t^e_i(\bm{a})/n_e}
  {\alpha^e_i}
\bigg]^2
+ \left(\frac{1-n_e}{\delta n_e}\right)^2 + \sum_k \big( r_k^e \big)^2 \right),
\label{e.chi2}
\end{align}
where $e$ labels different experimental datasets with data points $d^{\, e}_i$ and corresponding theory values $t_i^e(\bm{a})$, $\alpha_i^e$ are point-to-point uncorrelated uncertainties added in quadrature, and $\beta^{\, e}_{k,i}$ are the corresponding point-to-point correlated uncertainties. 
We allow the theory to be shifted additively by an amount
    $\sum_k r^e_k\, \beta^{\, e}_{k,i}$,
with nuisance parameters that are optimized via $\diff \chi^2/\diff r_k^e=0$.

In addition to fitting the shape parameters in Eq.~(\ref{e.PDF}), we also fit the overall multiplicative normalization factors $n_e$ in \eref{chi2} for all 7 datasets used in this analysis, with a Gaussian penalty controlled by the quoted experimental normalization uncertainties $\delta n_e$.
The total number of free parameters is therefore 15, comprising the 7 PDF shape parameters and 1 pion splitting function parameter for the LN data, and the 7 data normalizations (3~for the $\pT$-integrated E615 and NA10 194~GeV and 286~GeV data, 1 each for the $x_F$ and $Q$-integrated E615 data, and 1 each for the H1 and ZEUS LN datasets).

For the uncertainty quantification for generic observables ${\cal O}$ such as the PDFs or cross sections, we evaluate the expectation values ${\rm E}[\mathcal{O}]$ and variances ${\rm V}[\mathcal{O}]$ according to 
\begin{subequations}
\label{e.EV}
\begin{align}
{\rm E}[\mathcal{O}] 
  &= \frac{1}{N}\sum_k \mathcal{O}(\bm{a}_k),
  \\
{\rm V}[\mathcal{O}] 
  &= \frac{1}{N} \sum_k \big[ \mathcal{O}(\bm{a}_k) - {\rm E}[\mathcal{O}]\big]^2,
\end{align}
\end{subequations}
where $\{\bm{a}_k\}$ are the ensemble of Monte Carlo parameters of size $N$ drawn from the posterior distribution (\ref{e.Bayes}).
This definition avoids assuming Gaussianity for the parameter distributions, as in the traditional Hessian method~\cite{Pumplin:2002vw, Martin:2002aw, Watt:2012tq}.

The Monte Carlo samples are constructed from \eref{Bayes} using data resampling and performing multiple $\chi^2$ minimizations.
The starting values of the parameters for each minimization are selected randomly from flat priors within the nonvanishing region of the prior, and each data point $d_i^{\, e}$ is shifted Gaussianly within the quoted uncorrelated uncertainties $\alpha_i^e$ added in quadrature. 
In practice, the ranges of the $a_i$ parameters that control the $x \to 0$ behavior are restricted to guarantee integrability of the pion PDFs for the valence and momentum sum rules.

\subsection{Multiple solutions}
\label{s.multiple}

As is the nature of Monte Carlo analyses, typically multiple solutions are obtained for many of the PDF parameters.
In order to discriminate between the various solutions, we employ a $k$-means clustering algorithm~\cite{Steinhaus57, Bock07}, similar to that used in the recent JAM19 analysis of proton PDFs~\cite{JAM19}.
Discriminating between the various clusters of solutions is then achieved on the basis of the total $\chi^2$ for each of the clusters, or eliminating clusters that build up at edges of certain parameters with physically motivated boundaries.

Initially, fits were performed with the ranges of the $b_i$ parameters for the gluon and sea quark PDFs restricted so that these distributions do not exceed the valence quark PDF at large $x$.
After identifying the cluster with the smallest $\chi^2$, optimal priors were generated from the replicas in that cluster.
Each of the parameters in the priors was generated in a random normal distribution with the mean and standard deviation of the best cluster's parameters.
Solutions were removed in which the $b_s$ parameter for the sea quark PDF was smaller than the valence quark $b_v$ parameter, to ensure the natural dominance of the valence distribution at large $x$.
Cuts were also made to avoid edge effects in parameter space from the artificial buildup of replicas at the boundaries of the parameter ranges associated with the gradient descent algorithm.

\begin{figure}
\includegraphics[width=\textwidth]{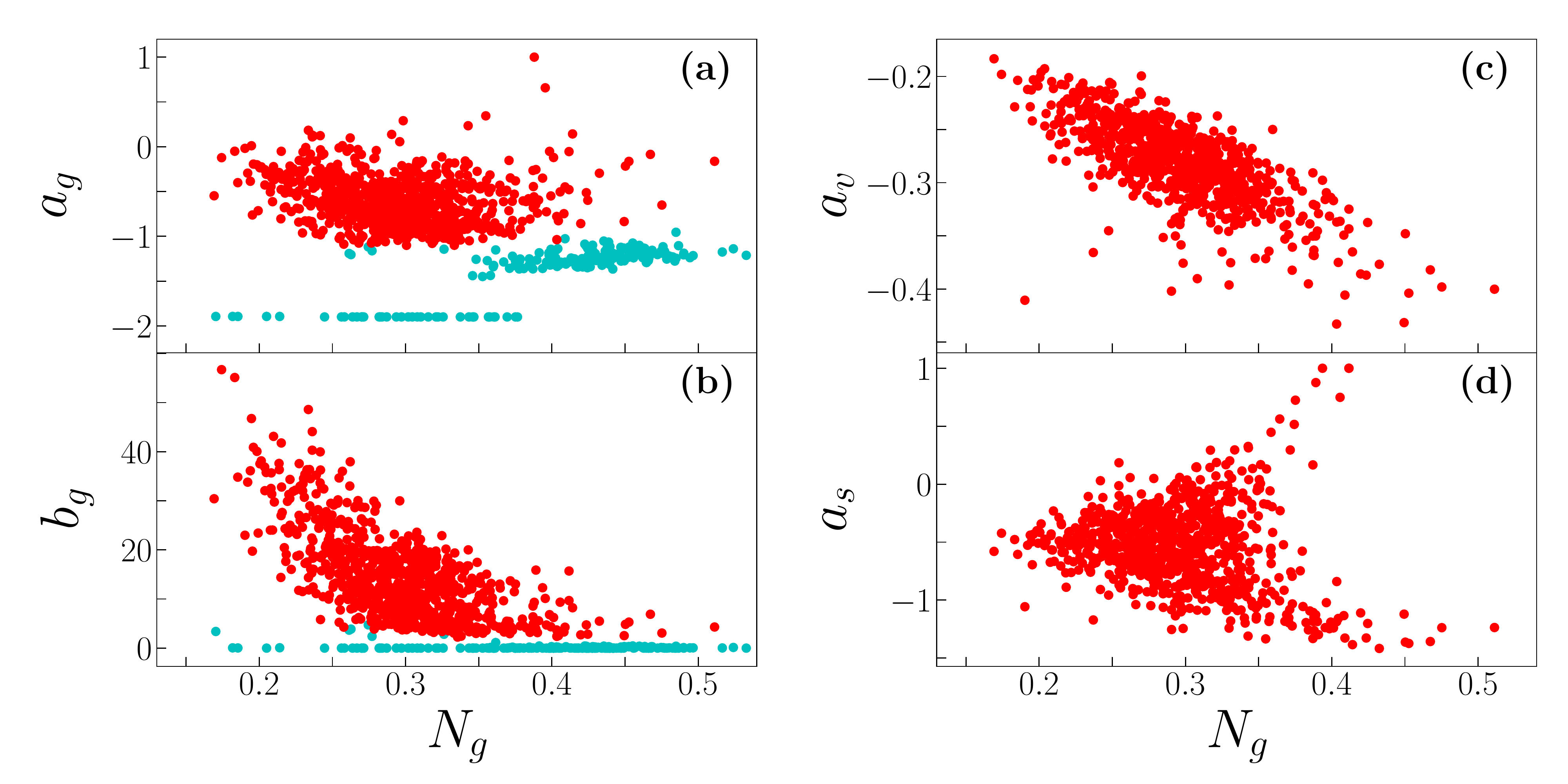}
\caption{Correlations between PDF parameter distributions for
    {\bf (a)} gluon $a_g$ parameter versus the gluon normalization $N_g$, and
    {\bf (b)} gluon $b_g$ parameter versus $N_g$, 
    including solutions that build up at the edges of parameter space (cyan),
    {\bf (c)} valence parameter $a_v$ versus $N_g$, and
    {\bf (d)} sea parameter $a_s$ versus $N_g$, after removing the cyan solutions.
    Note the ellipsoidal shape of the parameter distribution in {\bf (c)} suggests a Gaussian distribution, while the irregular shape in {\bf (d)} suggests a non-Gaussian 
    parameter distribution.}
\label{f.parameters}
\end{figure}

The spurious buildup at the boundaries is illustrated in \fref{parameters}(a) and (b), where the distribution of the gluon $a_g$ and $b_g$ parameters is shown versus the gluon normalization parameter $N_g$ in scatterplot form.
The $k$-means algorithm was then used to separate clusters of solutions based on these parameters.
A buildup of solutions is seen at the lower boundaries of both the $a_g$ and $b_g$ parameters, with the small-$x$ parameter $a_g > -2$ constrained to ensure a finite integrated momentum fraction carried by gluons, and $b_g > 0$ to avoid a diverging gluon PDF as $x \to 1$.
The solutions corresponding to parameters displaying obvious edge effects should not be regarded as true solutions, and merely reflect of the imperfections of the Monte Carlo algorithm adopted.

A virtue of the Monte Carlo analysis is that it allows us to avoid having to assume Gaussian distributions for the parameters.
As shown \fref{parameters}(c) for the scatterplot of the valence $a_v$ parameter versus the gluon normalization parameter $N_g$, after removing the accumulated solutions on the boundaries the remaining solutions are distributed in an approximately ellipsoidal shape, suggesting that these parameters indeed follow Gaussian distributions.
On the other hand, the scatterplot of the gluon $N_g$ versus the sea $a_s$ parameter in  \fref{parameters}(d) displays a more irregular shape, with the parameters not displaying Gaussianity.
In performing our Monte Carlo analysis, we admit all solutions, regardless of the shapes of their parameter distributions.
With new sufficiently precise data, one may in the future be able to remove the boundary constraints on the parameters and obtain the sum rules and valence versus sea ordering from the data alone.

\section{QCD Bayesian Analysis}
\label{s.results}

Having outlined the theoretical framework for our analysis, along with the datasets used and methodology employed to fit them, in this section we present the results of the fits, including data-to-theory comparisons and the resulting pion PDFs together with their uncertainties.
We begin the presentation of the results with a discussion of the dataset selection and justification of the cuts employed to ensure the integrity of our theoretical treatment of the data.

\subsection{Data selection}
\label{ssec.quality}

As discussed in Sec.~\ref{s.Datasets}, for the $\pT$-integrated Drell-Yan lepton-pair production data, we implement mostly the same kinematic cuts as in the previous JAM analysis~\cite{Barry18}, namely, $0 < x_F < 0.9$ and a slightly modified range $4.16 < Q < 7.68$~GeV, in order to avoid edges of phase space and ensure dominance of leading power factorization.
Similarly, for the HERA LN data, to restrict the analysis to the region where the pion-exchange process dominates the neutron production, we apply the cut $x_L > 0.8$, as in the earlier studies~\cite{McKenney, Barry18}.

\begin{figure}[t]
\centering
\includegraphics[width=0.65\textwidth]{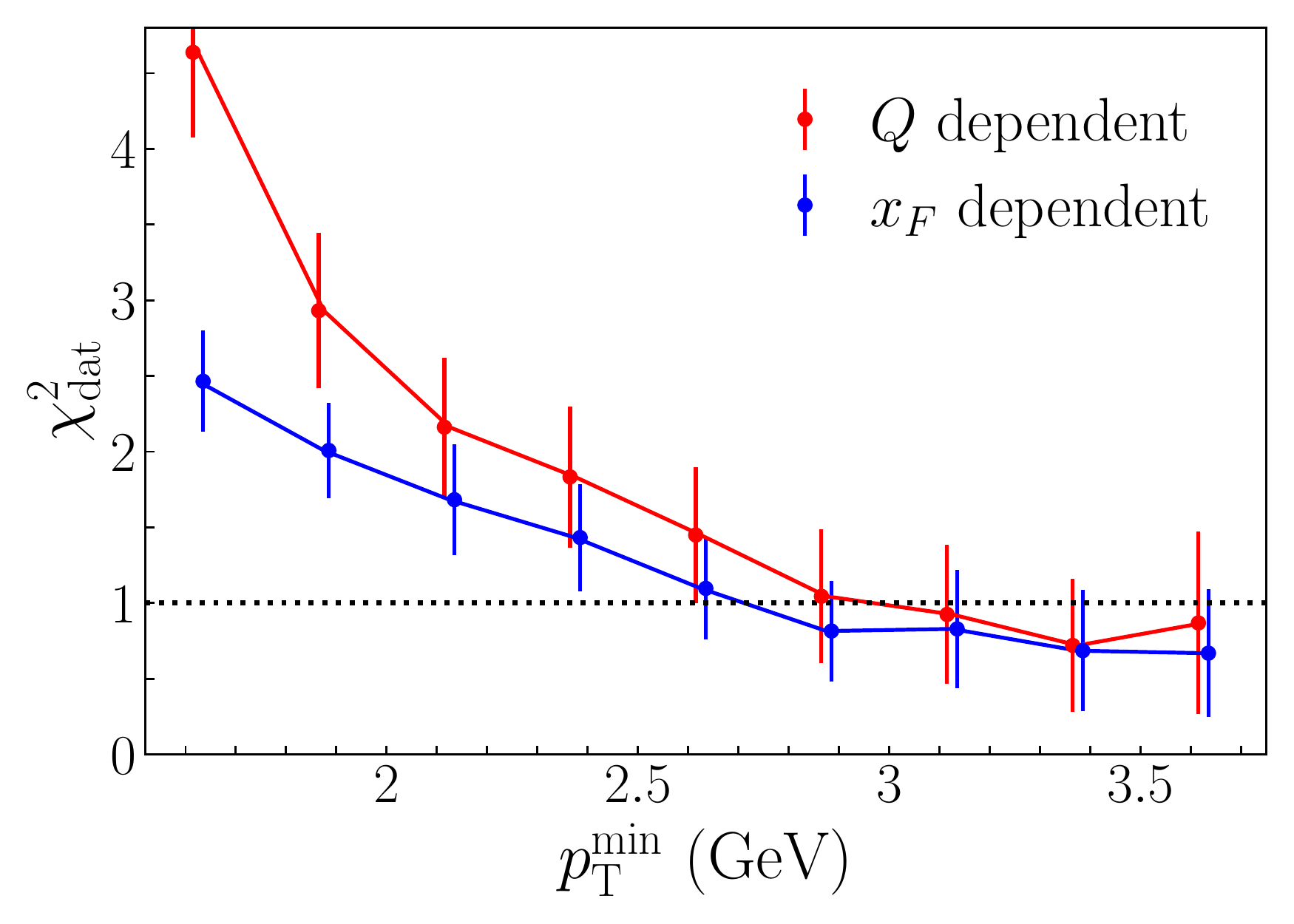}
\vspace*{-0.5cm}
\caption{Dependence of the $\chi^2_{\rm dat}$ values of the global fit on the minimum value of $\pT$, $p_{\rm T}^{\rm min}$, for the $Q$-dependent (red circles) and $x_F$-dependent (blue circles) $\pT$-differential Drell-Yan E615 data~\cite{E615} used in this analysis.}
\label{f.LPPpTcuts}
\end{figure}

For the new $\pT$-dependent Drell-Yan data considered in this analysis, the expectation is that the fixed-order term in Eq.~(\ref{e.w+y}) should be the dominant contribution at sufficiently large values of $\pT$ relative to the mass of the virtual photon.
Conversely, at small values of $\pT$ we expect the collinear approximation to break down, and a description in terms of TMD distributions to be more appropriate.
To test the range in $\pT$ down to which the fixed order calculation can be extended, we perform several analyses with varying values of $p_{\rm T}^{\rm min}$.
The results, shown in \fref{LPPpTcuts}, indicate that good fits can be obtained to the $\pT$-dependent Drell-Yan data (both $Q$ and $x_F$ dependent), with $\chi^2$ per datum of
$\chi^2_{\rm dat} \sim 1$, for $p_{\rm T}^{\rm min} > 2.7$~GeV.
Below this value the agreement with data deteriorates significantly, with $\chi^2_{\rm dat} \gtrsim 2-3$ for $p_{\rm T}^{\rm min} \lesssim 2$~GeV.
On the other hand, increasing the minimum $\pT$ to larger values does not improve the $\chi^2_{\rm dat}$ further, while at the same time reducing the number of points included in the fit (from 83 points with $p_{\rm T}^{\rm min} = 2.7$~GeV down to 39 points with $p_{\rm T}^{\rm min} = 3.5$~GeV).
To maximize the number of data points to be fitted, without disrupting the integrity of the fit, we choose a uniform cut for both the $Q$- and $x_F$-dependent datasets of $p_{\rm T}^{\rm min} = 2.7$~GeV.

\subsection{Data versus theory comparison}
\label{ssec.datatheory}

\begin{table}[b]
\centering
\caption{Summary of results for the global fit to the Drell-Yan (DY) $\pT$-integrated and $\pT$-differential data, and the leading neutron production structure functions and ratios from HERA, including the number of data points fitted, $N_{\rm dat}$, the data normalization factors, $n_e$ [Eq.~(\ref{e.chi2})], and the $\chi^2_{\rm dat}$ values. The DY$\pT$ data were fitted with the scale set to $\mu=\pT/2$ for the PDFs.\\}
\begin{tabular}{llccc}
\hline
~Process~~ & ~Experiment (observables)~~  
        & ~~~$N_{\rm dat}$~~~  
        & ~~~$\chi^2_{\rm dat}$~~~   
        & ~~~$n_e$~~~ \\
\hline
{\bf ~DY}& ~E615 $(x_F,Q)$          & 61 
                                    & 0.85 & 1.08 \\
         & ~NA10~\mbox{\scriptsize (194~GeV)} $(x_F,Q)$
                                    & 36 
                                    & 0.52 & 0.88 \\
         & ~NA10~\mbox{\scriptsize (286~GeV)} $(x_F,Q)$ 
                                    & 20 
                                    & 0.78 & 0.83 \\
         & ~E615 $(Q,\pT)$          & 34 
                                    & 1.08 & 0.83 \\
         & ~E615 $(x_F,\pT)$        & 49 
                                    & 0.85 & 0.50 \\ \hline
{\bf ~LN}& ~H1                      & 58 
                                    & 0.38 & 1.26 \\
         & ~ZEUS                    & 50 
                                    & 1.51 & 0.95 \\ 
\hline
~\textbf{Total} &                   & {\bf 308} 
                                    & {\bf 0.85} &   \\ \hline \\
\end{tabular}
\label{t.chi2}
\end{table}

With these cuts on the data, the results of the global fit are summarized in \tref{chi2}, where for each experiment we list the number of data points fitted, the data normalization parameter, and the corresponding $\chi^2_{\rm dat}$ value.
The overall $\chi^2$ is 262 for a total of 308 data points, with the total $\chi^2_{\rm dat} = 0.85$ per datum.
Moreover, we find that fairly good fits are obtained to all of the datasets, with $\chi^2_{\rm dat} \approx 1$ for each experiment.
An exception is the ZEUS LN dataset, with a $\chi^2_{\rm dat} \approx 1.5$, which partly reflects the relatively small uncertainties on these data compared with the H1 results.

\begin{figure}[t]
\centering
\includegraphics[width=0.75\textwidth]{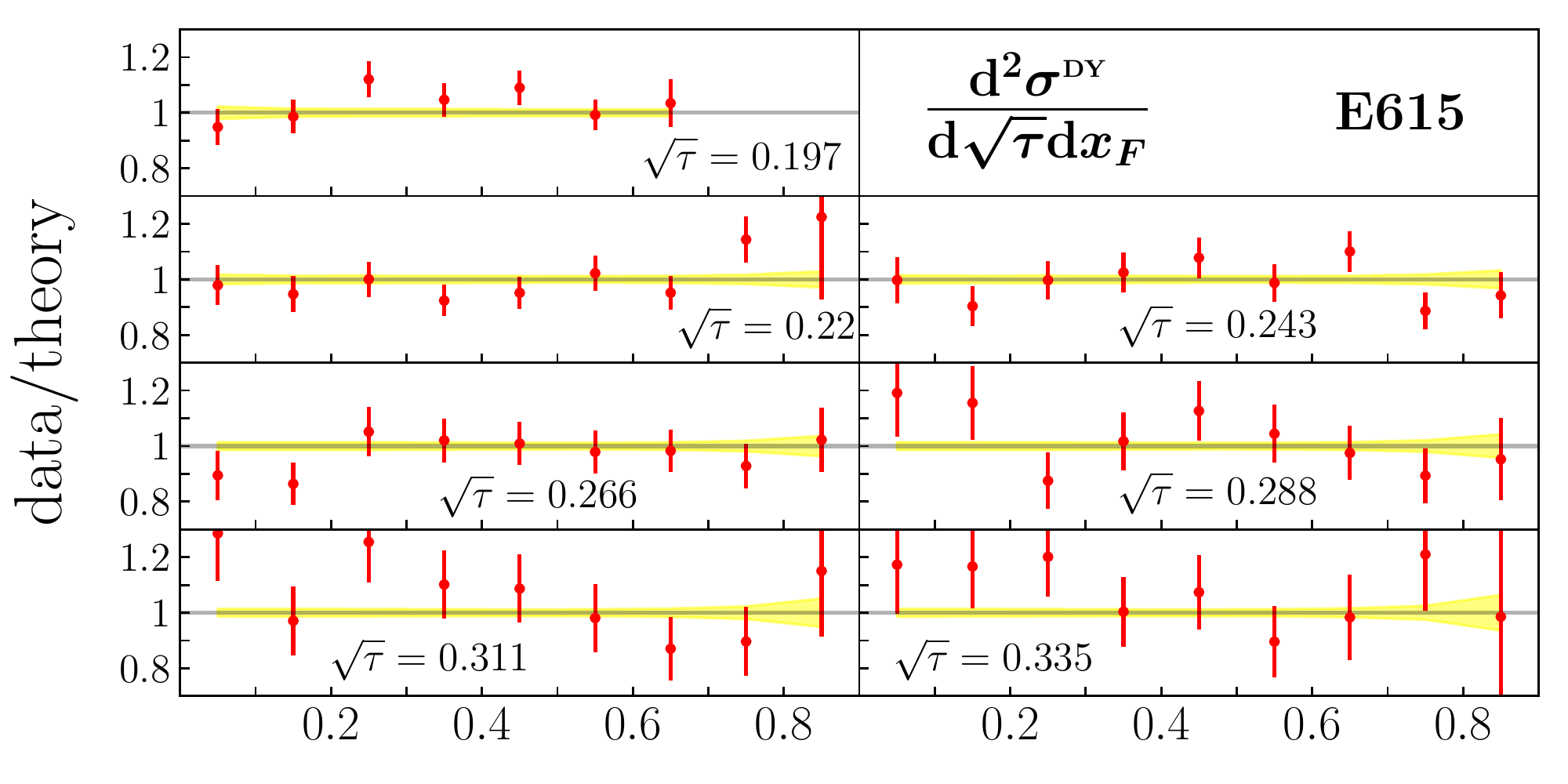}
\includegraphics[width=0.75\textwidth]{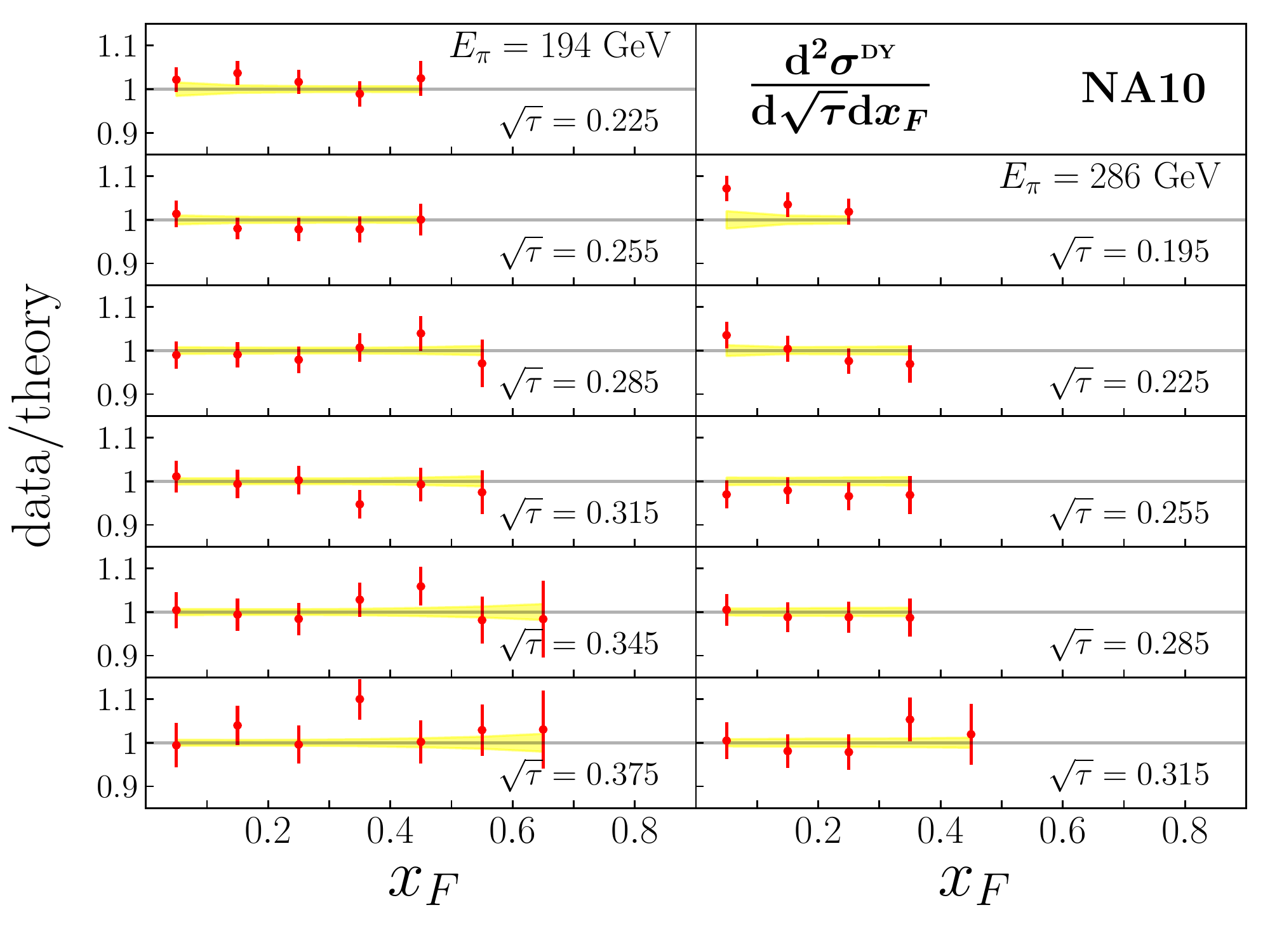}
\vspace*{-0.5cm}
\caption{Data-to-theory ratios for the $x_F$ dependence of the Drell-Yan cross section  $\diff^2\sigma^{\mbox{\tiny \rm DY}}/\diff \sqrt{\tau} \diff x_F$ at fixed values of $\sqrt{\tau}$ from the E615~\cite{E615} {\bf (top)} and NA10~\cite{NA10} {\bf (bottom)} experiments. The NA10 data are separated for the two pion beam energies of 194~GeV {\bf (bottom left)} and 286~GeV {\bf (bottom right)}, and the yellow bands represent the uncertainty on the theory calculations.}
\label{f.DYdot}
\end{figure}

\begin{figure}[t]
\centering
\includegraphics[width=0.56\textwidth]{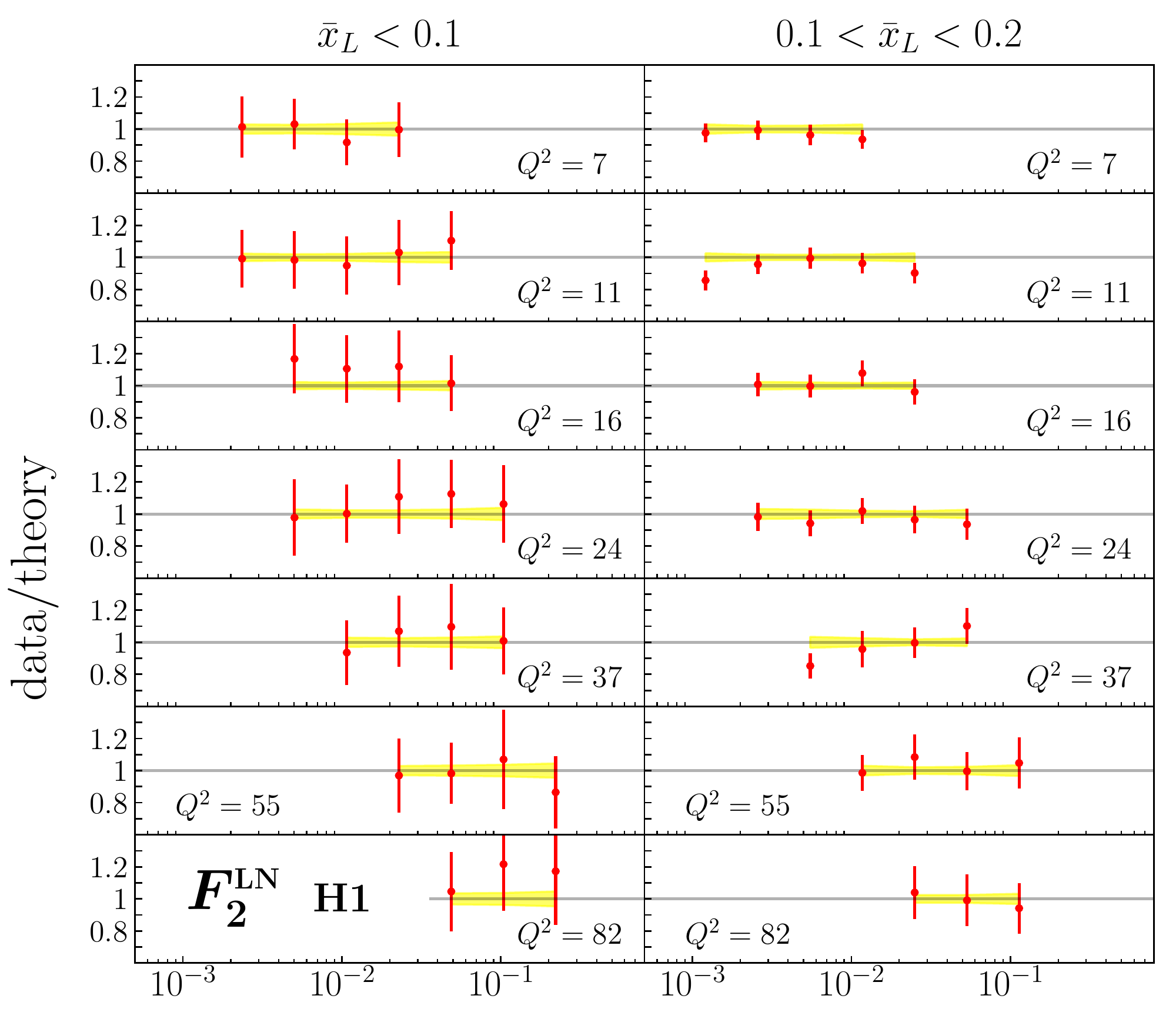}
\includegraphics[width=0.56\textwidth]{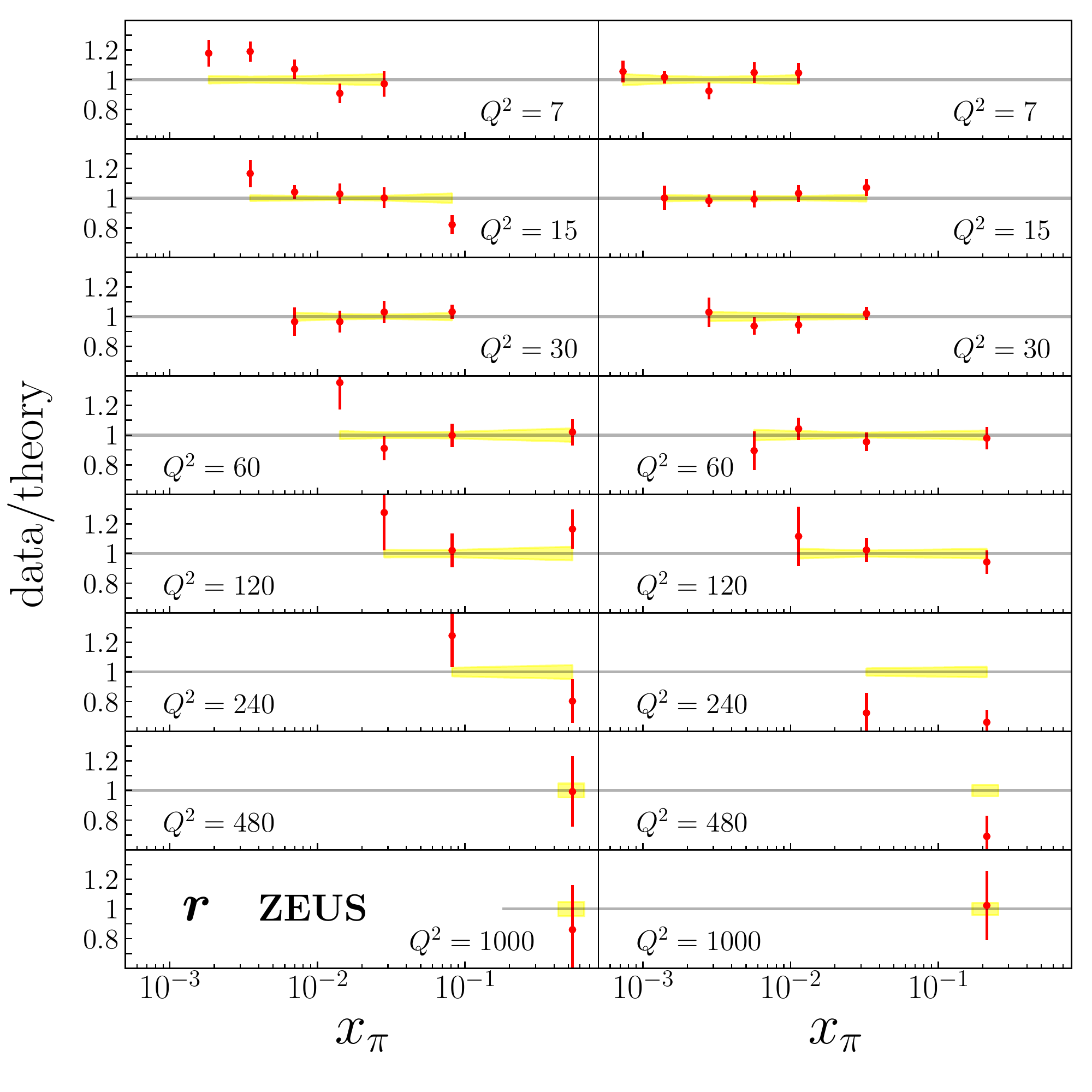}
\vspace*{-0.5cm}
\caption{Data-to-theory ratios for the $x_\pi$ dependence of the leading neutron electroproduction data for the $F_2^{\LN}$ structure function from H1~\cite{H1} [Eq.~(\ref{eq.LNxsec})] {\bf (top)} and the structure function ratio $r$ from ZEUS~\cite{ZEUS} [Eq.~(\ref{e.ZEUSrxsec})] {\bf (bottom)}, in bins of fixed $Q^2$ (in GeV$^2$). The yellow band represents the theoretical uncertainties.}
\label{f.LNdot}
\end{figure}

\begin{figure}[t]
\centering
\includegraphics[width=0.7\textwidth]{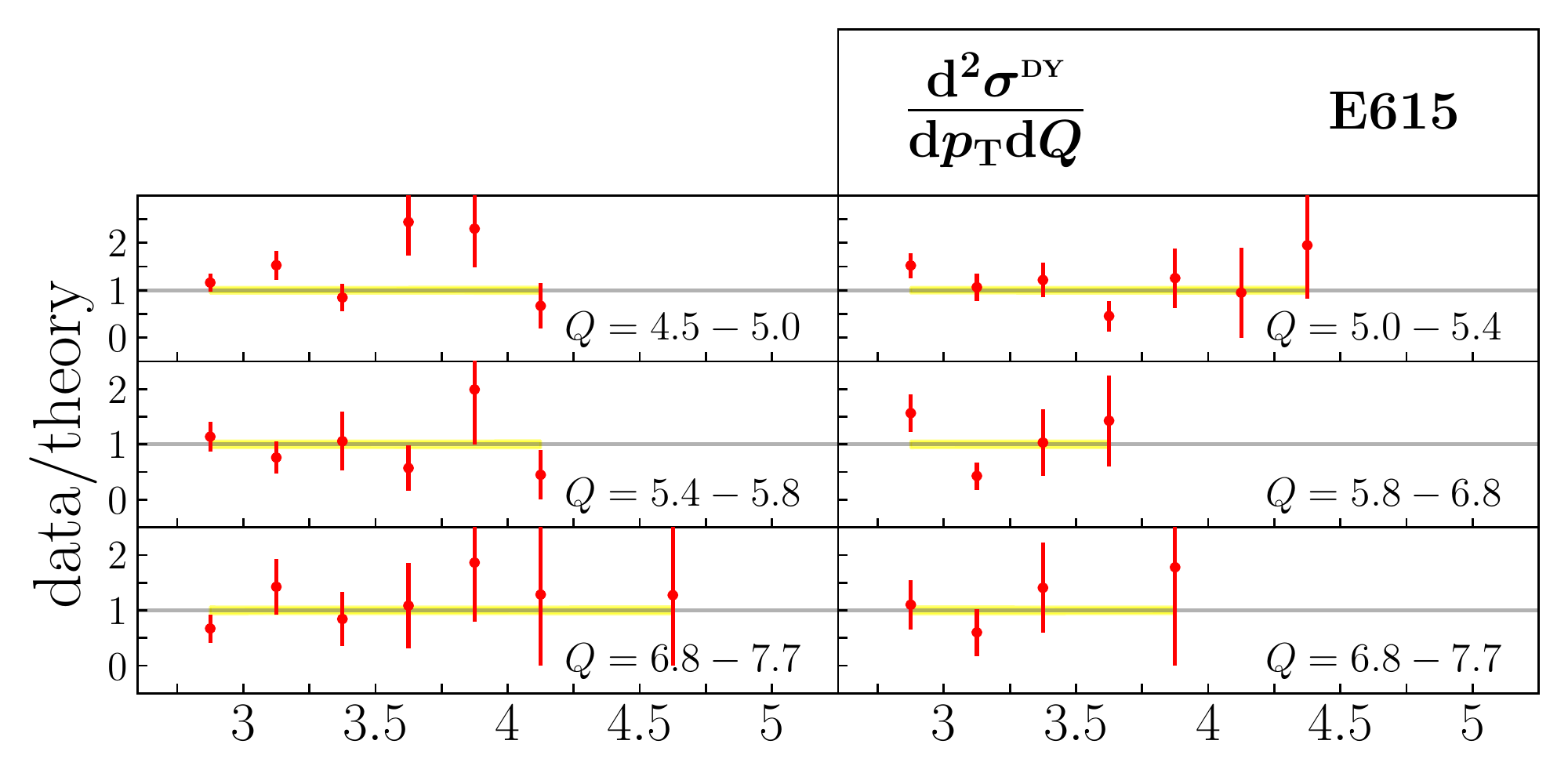}
\includegraphics[width=0.7\textwidth]{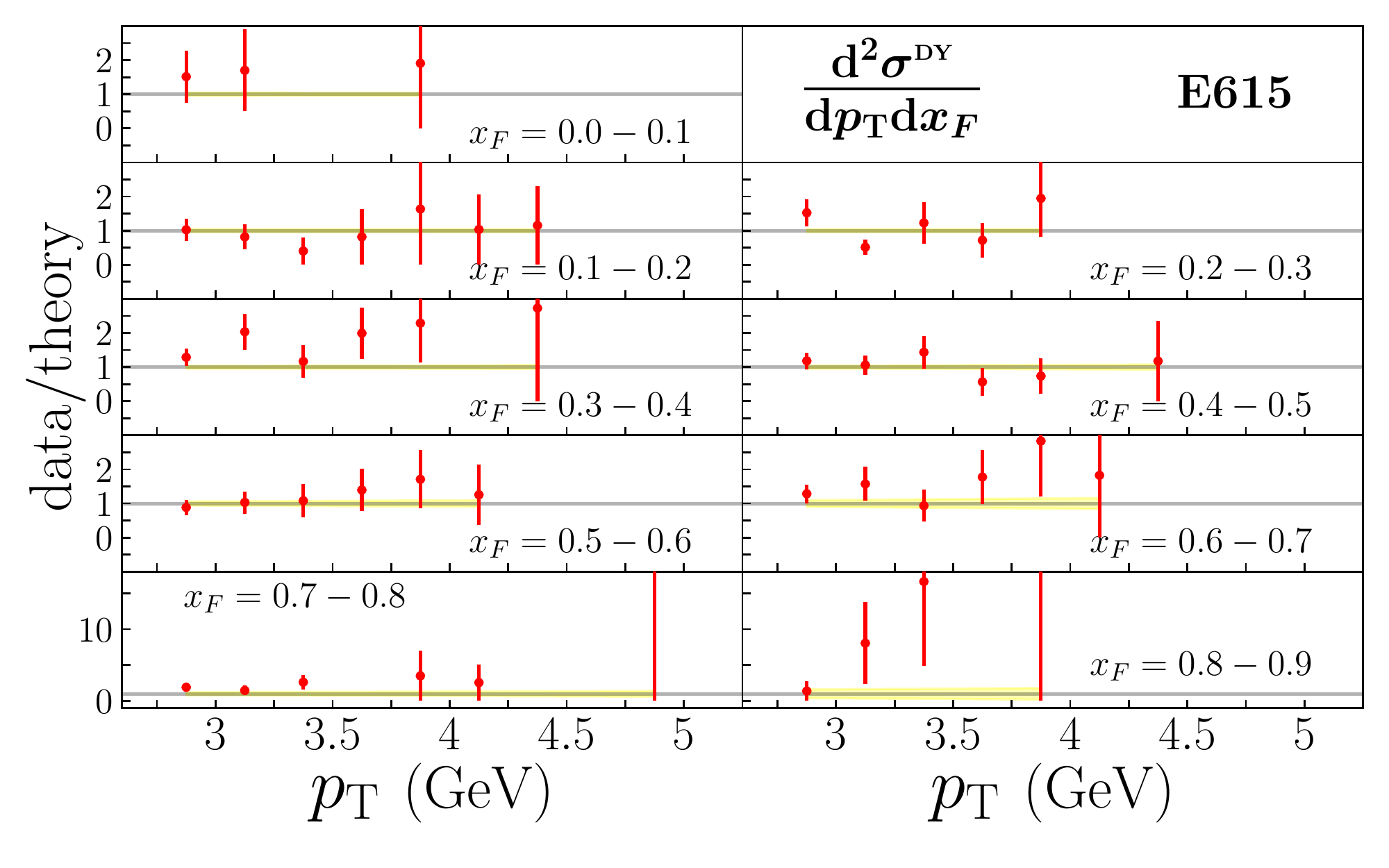}
\vspace*{-0.5cm}
\caption{Data-to-theory ratios for the $\pT$ dependence of the Drell-Yan cross sections $\diff^2\sigma^{\mbox{\tiny \rm DY}}/\diff \pT \diff Q$ {\bf (top)} and $\diff^2\sigma^{\mbox{\tiny \rm DY}}/\diff \pT \diff x_F$ {\bf (bottom)} from the E615 experiment~\cite{E615}. The yellow bands represent the theoretical uncertainties, and the scale is set to $\mu=p_T/2$ for the PDFs.}
\label{f.DYpTdot}
\end{figure}

The comparison between theory and experiment is illustrated in Figs.~\ref{f.DYdot} -- \ref{f.DYpTdot}, which show data over theory ratios for the $\pT$-integrated Drell-Yan data from E615~\cite{E615} and NA10~\cite{NA10}, the leading neutron electroproduction data from H1~\cite{H1} and ZEUS~\cite{ZEUS}, and the $\pT$-differential Drell-Yan data not considered in previous analyses.
For each of the theoretical observables, as defined in Sec.~\ref{sec.Theory}, the cross sections calculated in terms of the fitted PDFs are divided by the $n_e$ normalization parameter, and allowed to be shifted by an amount given by the nuisance parameters, as in Eq.~(\ref{e.chi2}).
The values compared with individual data points are a calculated mean over all of the replicas from the Monte Carlo fits, with the mean and variance in Figs.~\ref{f.DYdot} -- \ref{f.DYpTdot} computed as in Eqs.~(\ref{e.EV}).

The comparison with the Drell-Yan cross sections $\diff^2 \sigma^{\mbox{\tiny \rm DY}}/\diff \sqrt{\tau} \diff x_F$ in \fref{DYdot} indicates that the data can be well described by the fitted pion PDFs within the framework of the perturbative QCD calculation at next-to-leading order (NLO) in $\alpha_s$.
The data-to-theory ratios are shown as a function of $x_F$ in various bins of $\sqrt{\tau}$ for both the Fermilab E615~\cite{E615} and CERN NA10~\cite{NA10} data sets, with the latter separated into the two pion beam energies, $E_\pi=194$ and 286~GeV.
The ratios are generally consistent with unity, within the uncertainties of the data, across the entire range of $x_F$ and $\sqrt{\tau}$ shown, with $\chi^2_{\rm dat}$ values $\lesssim 1$ for both experiments.
The experimental uncertainties on the NA10 data are somewhat smaller than the uncertainties on the E615 data, although the E615 data extended to larger values of $x_F$.
The theory uncertainty bands indicated in the ratios reflect the uncertainties on the PDFs, which increase at the highest values of $x_F$.

For the comparisons with the LN 
data from HERA, in \fref{LNdot} we show the data-to-theory ratios of the $F_2^{\LN}$ structure function [Eq.~(\ref{eq.LNxsec})] from H1~\cite{H1} and the ratio $r$ [Eq.~(\ref{e.ZEUSrxsec})] of the leading neutron to inclusive proton cross sections from ZEUS~\cite{ZEUS}.
The ratios are shown as a function of $x_\pi$ over a large range of $Q^2$ bins, ranging from $Q^2=7$~GeV$^2$ to $Q^2=1000$~GeV$^2$, for two bins of momentum fraction $\bar{x}_L$ carried by the exchanged charged particle (pion), restricted to $\bar{x}_L < 0.1$ and $0.1 < \bar{x}_L < 0.2$ to ensure pion exchange dominance~\cite{McKenney, Barry18}.
Within the quoted uncertainties, the H1 data can be well described by our fit across all the kinematics shown, with a $\chi^2_{\rm dat}$ value of $\approx 0.4$ per datum.
The systematic uncertainties on the H1 data are largest for the lowest-$\bar{x}_L$ bin, and since the magnitude of the $F_2^{\LN}$ structure function increases with $\bar{x}_L$, the relative uncertainties for the data are thus largest for $\bar{x}_L<0.1$

For the ZEUS data, since these are presented as a ratio of semi-inclusive to inclusive structure functions, the uncertainties are smaller than for the absolute structure functions from the H1 experiment, with a number of the correlated systematic uncertainties canceling.
Consequently, these data provide the strongest constraints, and give the highest overall $\chi^2$ of all the datasets fitted in this analysis, with $\chi^2_{\rm dat} \approx 1.5$.
There is some tension with the data at the smallest $x_\pi$ values, for the lowest $Q^2$ bin, but generally the agreement between theory and data is quite good.

The $\pT$-differential Drell-Yan cross sections from E615~\cite{E615}, which includes the new data considered in our analysis, are displayed in \fref{DYpTdot} as data-to-theory ratios versus the $\pT$ of the lepton pair.
Choosing the QCD factorization to be $\mu=\pT/2$ (see Sec.~\ref{ssec.scale}), fairly good fits are obtained for both the $x_F$-integrated data ($\chi^2_{\rm dat}=1.08$) and the $Q$-integrated data ($\chi^2_{\rm dat}=0.85$) shown in \fref{DYpTdot}, although a somewhat small value of the overall data normalization factor is necessary for the $Q$-integrated ($n_e=0.50$) compared with the $x_F$-integrated ($n_e=0.83$) cross sections.
The difference between the normalization factors for the various datasets reflects possible tensions among the data, which can affect deviations of $n_e$ outside of the normalization uncertainties reported by the experiments.
The uncertainties on the $\pT$-dependent data are generally much larger than on the $\pT$-integrated cross sections in \fref{DYdot} (note the vertical scale on the data/theory ratios in \fref{DYpTdot}), and grow with $\pT$, increasing markedly as $x_F \to 1$.

Future, higher-precision data on the $\pT$ dependence of Drell-Yan lepton-pair production cross sections would provide stronger constraints on the fits than the currently available data.
A program of pion-induced and kaon-induced Drell-Yan experiments is being discussed in connection with the proposed COMPASS++/AMBER facility at CERN~\cite{Denisov:2018unj}.
Nevertheless, given the present difficulty in reconciling the $\pT$ dependence of proton-induced Drell-Yan data in the large-$\pT$ region with the latest theoretical tools~\cite{Bacchetta:2019tcu}, it is noteworthy that a consistent description of $\pT$-differential and $\pT$-integrated pion-nucleus cross sections can be achieved within the same collinear factorization framework.

\subsection{Parton distributions}
\label{ssec.fits}

\begin{figure}[t]
\centering
\includegraphics[width=0.8\textwidth]{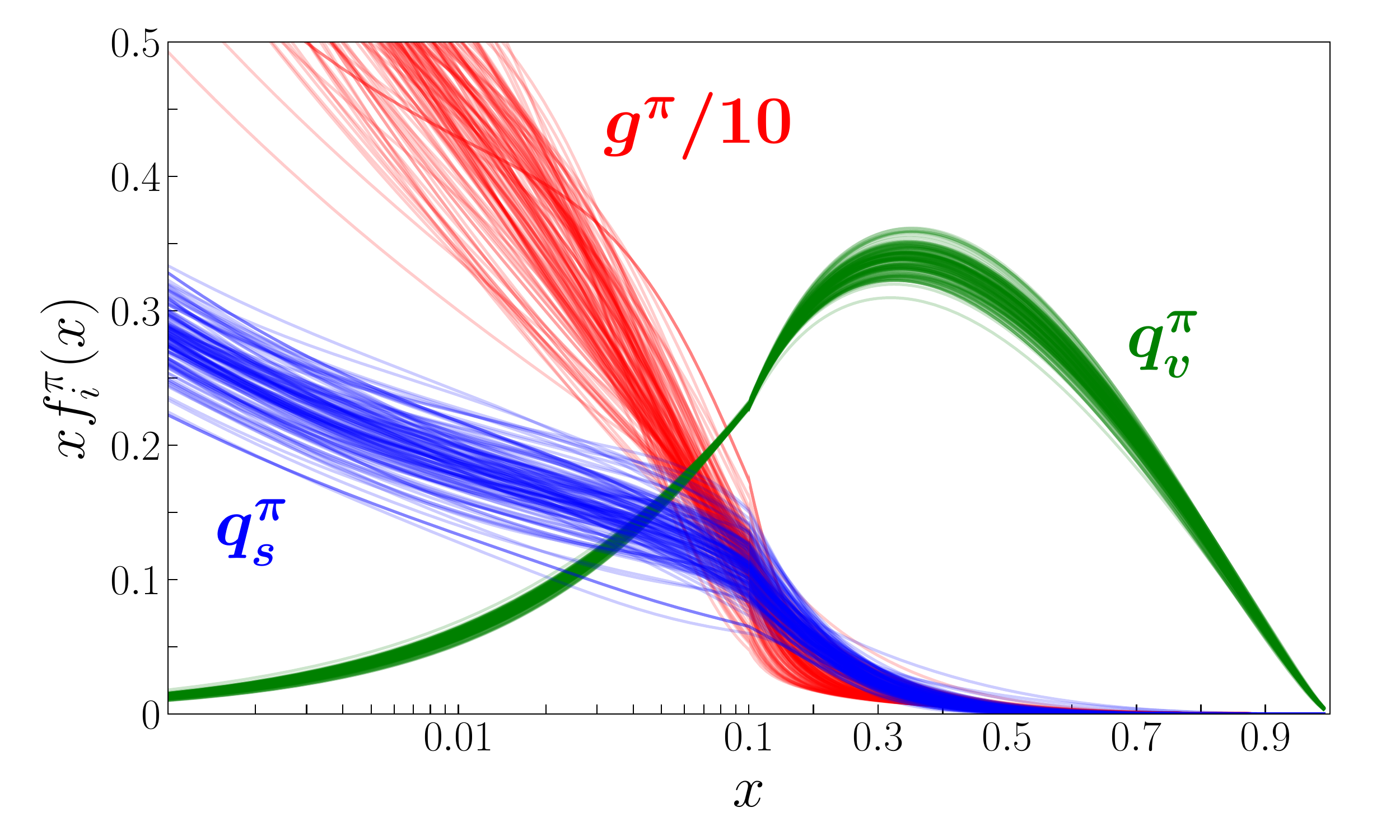}
\vspace*{-0.5cm}
\caption{Pion valence quark, sea quark and gluon (scaled by a factor 1/10) PDFs extracted from the JAM Monte Carlo analysis of the Drell-Yan ($\pT$-integrated and $\pT$-differential) and LN data at a scale of $\mu^2=10$~GeV$^2$.  Shown is a random sample of 150 replicas from the $\approx 800$ total number in our analysis.  Note that $x$ times the PDFs are shown, $xf_i^\pi(x)$.}
\label{f.PDFs}
\end{figure}

The resulting spectrum of PDFs from our Monte Carlo analysis of Drell-Yan ($\pT$-integrated and $\pT$-differential) and LN data, is illustrated in \fref{PDFs}, where we present the individual valence quark $q_v^\pi$, sea quark $q_s^\pi$, and gluon $g^\pi$ distributions (where $g^\pi$ is scaled by a factor 1/10) at a scale $\mu^2=10$~GeV$^2$.
For clarity, we show a random sample of 150 replicas out of the total $\approx 800$ replicas from our analysis.
Since some of the samples are faulty because of the imperfections of the Monte Carlo algorithm rather than viable physical solutions, the solutions displaying edge effects in Fig.~\ref{f.parameters} have been removed from the final sample.
The larger spread of solutions for sea quark and gluon PDFs compared with the valence quark distribution reflects the weaker constraints on the pion sea at small $x$ values from existing (mostly LN) data compared with those from the Drell-Yan data.
More data at low $x$ values would clearly be useful for constraining the sea quark and gluon PDFs.

A comparison of the JAM pion PDFs and their $1\sigma$ uncertainty bands with results from previous global analyses is shown in \fref{PDFcomps} at a scale $\mu^2=10$~GeV$^2$.
Generally good agreement, especially for the valence quark distribution, is found with the recent xFitter parametrization~\cite{xFitter20}, which fitted the $\pT$-integrated Drell-Yan data along with data on prompt photon production from the WA70 experiment~\cite{WA70}.
At low $x$ values the xFitter sea quark PDF has somewhat larger uncertainty, which reflects the fact no LN data were used in this fit, hence the diminished constraining power for the PDFs in this region.
For reference, the older GRV parametrization~\cite{GRV}, which preceded the HERA data and does not have PDF uncertainties, is also shown.
Compared to the JAM result, the GRV fit has a slightly softer valence PDF, $\sim 25\%$ smaller at intermediate $x \sim 0.3-0.5$, which is compensated by a harder gluon distribution compared with the JAM result.

\begin{figure}[t]
\centering
\includegraphics[width=\textwidth]{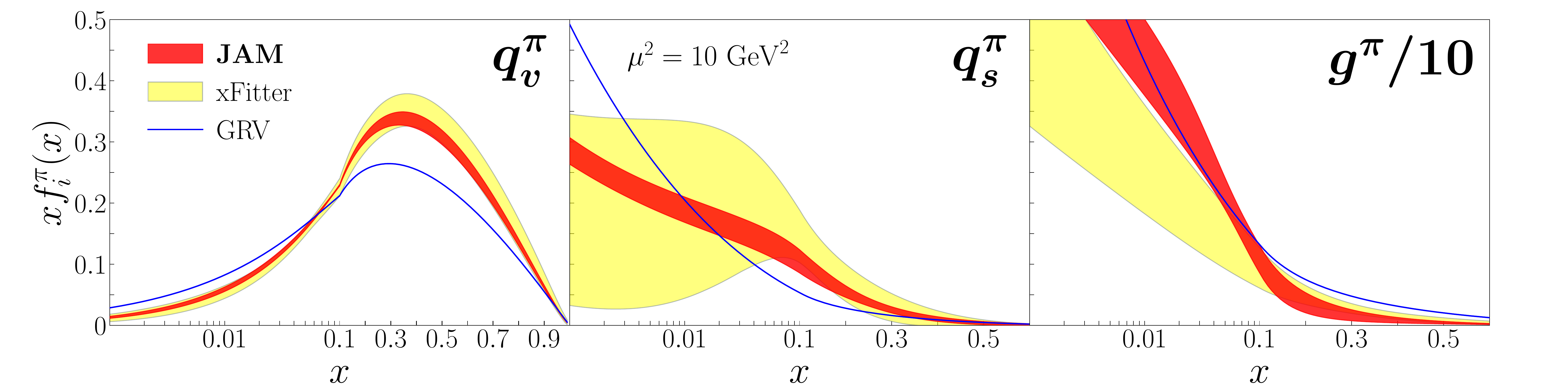}
\vspace*{-1cm}
\caption{Comparison of the pion valence quark $q_v^\pi$, sea quark $q_s^\pi$, and gluon $g^\pi$ (scaled by 1/10) PDFs from the current JAM analysis (red bands) at $\mu^2=10$~GeV$^2$ with the xFitter results~\cite{xFitter20} (yellow bands) and the GRV parametrization~\cite{GRV} (blue lines). The uncertainty bands represent $1\sigma$ CL.}
\label{f.PDFcomps}
\end{figure}

\begin{figure}[t]
\centering
\includegraphics[width=0.95\textwidth]{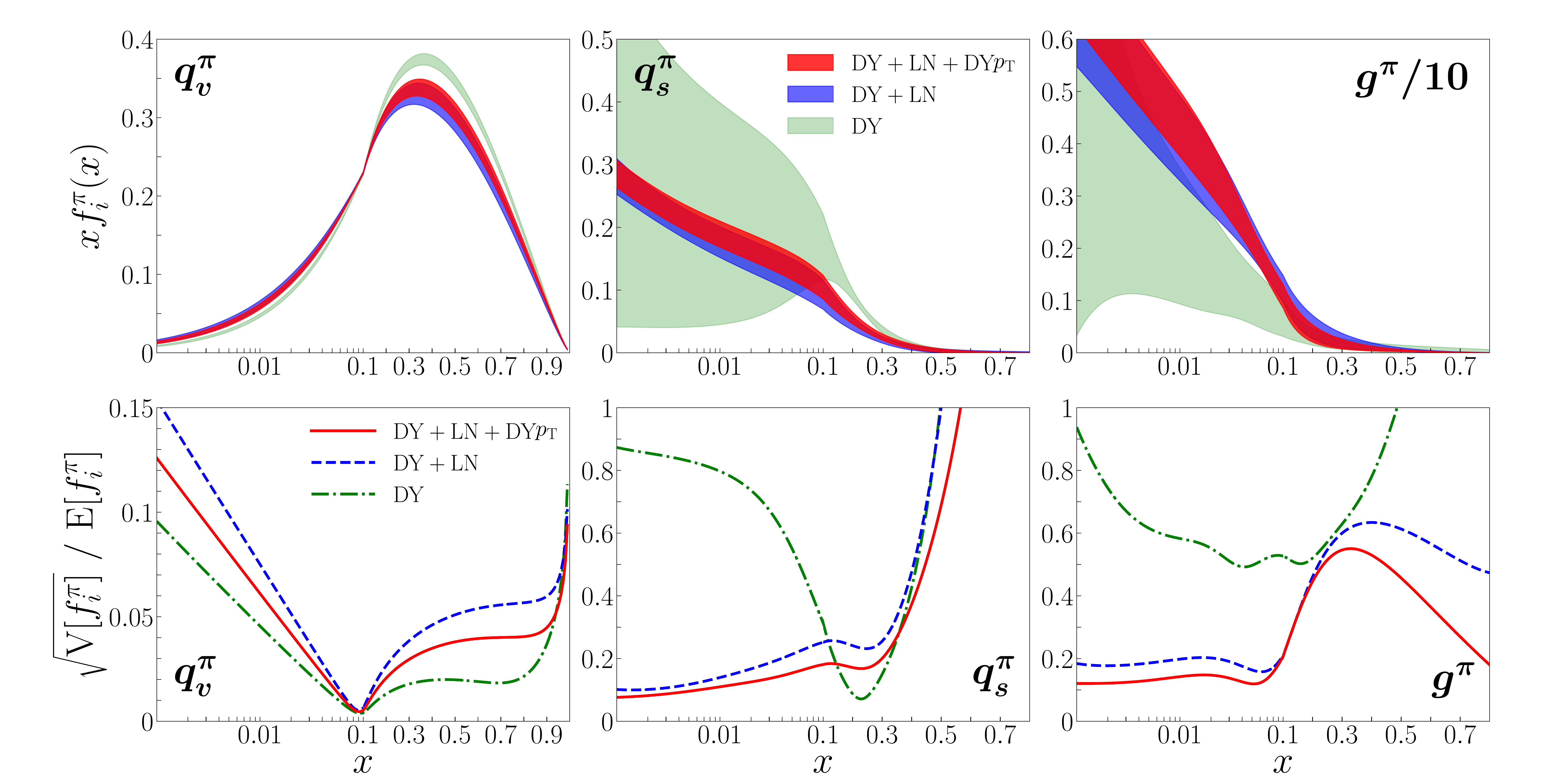}
\caption{Impact of datasets on pion valence quark {\bf (left)}, sea quark {\bf (middle)} and gluon {\bf(right)} PDFs at $\mu^2$=10~GeV$^2$.
    {\bf (Top)} Reduction of PDF $x f_i^\pi$ uncertainty bands from fitting only $\pT$-integrated Drell-Yan data (green), Drell-Yan and LN (blue), and $\pT$-integrated and $\pT$-dependent Drell-Yan and LN data (red).
    {\bf (Bottom)} Corresponding relative 1$\sigma$ uncertainties, as ratios of the square roots of the variances divided by the expectation values $\sqrt{{\rm V}}/{\rm E}$ for each PDF flavor $f_i^\pi$, for each of the datasets fitted.}
\label{f.PDFuncertainties}
\end{figure}

The impact of each of the datasets used in our analysis on the extraction of the pion PDFs is indicated in Fig.~\ref{f.PDFuncertainties} at a scale $\mu^2=10$~GeV$^2$.
In particular, Monte Carlo sampling has been carried out for three data selections: 
  {\bf (i)} $\pT$-integrated Drell-Yan only;
  {\bf (ii)} $\pT$-integrated Drell-Yan and LN data; and
  {\bf (iii)} $\pT$-integrated and $\pT$-differential Drell-Yan along with with LN data.
The effects on the PDFs and their $1\sigma$ uncertainties of adding each new dataset sequentially is shown, together with the relative errors with respect to the mean values of each data selection fit, as ratios of the square roots of the variances divided by the expectation values,
    $\sqrt{{\rm V}[f_i^\pi]}/{\rm E}[f_i^\pi]$.
While data selection in scenario (i) allows reasonably tight constraints on the valence quark PDF $q^\pi_v$, the sea quark $q^\pi_s$ and gluon $g^\pi$ PDFs have very large uncertainties.
Clearly the biggest overall impact on the PDFs uncertainties is scenario (ii), in which the addition of the HERA LN data constrains significantly the small-$x$ region for the gluon and the sea distributions, with modest effect on the valence distribution.
This is consistent with what was previously observed in Ref.~\cite{Barry18}.

The novel addition of the $\pT$-dependent Drell-Yan data in scenario (iii), has a modest impact on the shapes of the pion PDFs and their uncertainties.
The strongest impact is on the gluon distribution at large values of $x$, $x \gtrsim 0.3$.
This may be expected, given the sensitivity of the $\pT$-differential cross section on the pion's gluon PDF at lowest order in $\alpha_s$.
However, since the cross section at large $x$ is still mostly dominated by contributions from valence quarks, the overall impact on the glue is not overwhelming.
In other kinematic regions, the reduction in the PDF uncertainties after inclusion of the $\pT$-dependent Drell-Yan data is also relatively small, which reflects the larger errors of these data in \fref{DYpTdot} than for the $\pT$-integrated Drell-Yan and LN data in Figs.~\ref{f.DYdot} and \ref{f.LNdot}, respectively.

Interestingly, the behavior of the valence PDF at large $x$ is consistent with a $\sim (1-x)$ shape, as was found in Ref.~\cite{Barry18}. 
The $\pT$-dependent Drell-Yan does not alter this conclusion.
Further discussion on the large-$x$ behavior of the pion PDF in the presence of threshold resummation will be discussed elsewhere~\cite{PionResum}.

\begin{figure}[t]
\centering
\includegraphics[width=0.8\textwidth]{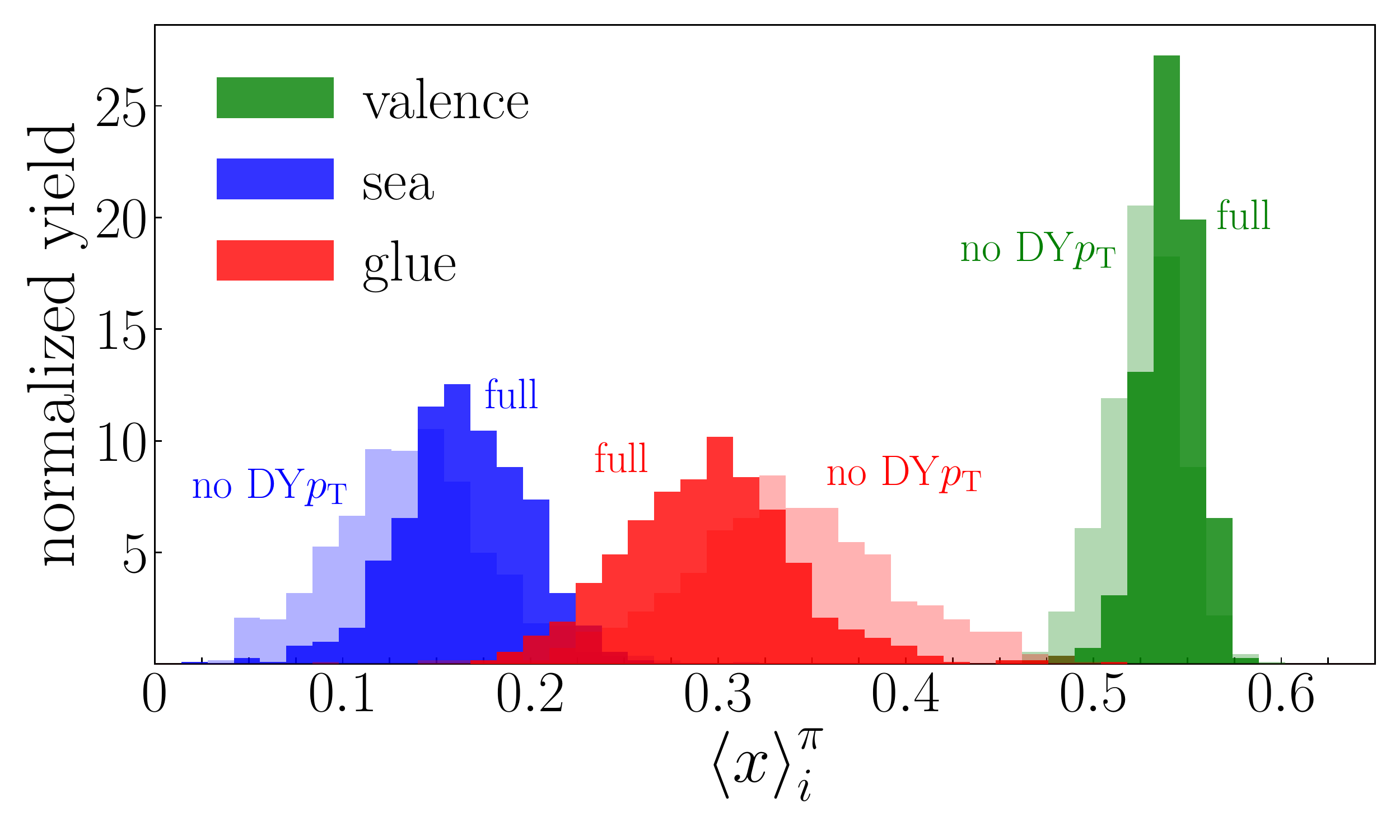}
\vspace*{-0.5cm}
\caption{Normalized yield of momentum fractions $\langle x \rangle^\pi_i$ of the pion carried by valence quarks (green), sea quarks (blue), and gluons (red) at 
$\mu^2 = m_c^2$, for the ``full'' analysis (darker shaded histograms) and without the $\pT$-dependent DY data (``no DY$\pT$'', lighter shaded histograms).}
\label{f.xpi}
\end{figure}

The impacts of the different datasets can be further explored by considering the momentum fractions carried by the individual flavors $i$ $(= v, s, g)$ of the pion, defined as
\begin{eqnarray}
\langle x \rangle_i^\pi &=& \int_0^1 \diff x\, x\, f_i^\pi(x,\mu^2),
\label{eq.xpi}
\end{eqnarray}
at a scale $\mu^2$.
The Monte Carlo samples for the pion momentum fractions are shown in Fig.~\ref{f.xpi} at the input scale $\mu^2=m_c^2$ as histograms, for each flavor, for the scenarios (ii) and (iii) described above, with the ``normalized yield'' defined by the area under the histogram being unity.
The momentum fractions with PDFs extracted from only the Drell-Yan data (scenario (i)) are not shown in Fig.~\ref{f.xpi} to avoid overcluttering, but their momentum fractions are similar to what was found previously in Ref.~\cite{Barry18}.
Namely, in this scenario the total quark valence momentum fraction is relatively well constrained to be 0.59(1), while the sea quark and gluon fractions, 0.28(10) and 0.13(11), respectively, have significantly larger uncertainties.

For the fit with the $\pT$-integrated Drell-Yan and LN data, labeled as ``no DY$\pT$'' in Fig.~\ref{f.xpi}, the inclusion of the HERA data reduces the uncertainty on the sea considerably.
The valence, sea quark and gluon momentum fractions in this case are 0.53(2), 0.14(4) and 0.34(6), respectively.
As observed in Ref.~\cite{Barry18}, the sea quark and gluon momentum fractions approximately switch in magnitude with the addition of the LN data, and the uncertainties on both decrease by a factor from $\sim 2$ to $\sim 4$.
With the better determined sea quark and gluon fractions, the valence quark fraction decreases by $\approx 10\%$ in order to satisfy the momentum sum rule.

\begin{table}[b]
\centering
\caption{Momentum fractions of the pion carried by valence quarks, sea quarks and gluons at the input scale, $\mu^2=m_c^2$, and at $\mu^2=10$~GeV$^2$, for various combinations of datasets used in this analysis. The results for the full analysis (``DY+LN+DY$\pT$'') are given in boldface.}
\begin{tabular}{l|lll|lll} \hline
&\multicolumn{3}{l|}{~~~~~~~~~~~~~~~$\mu^2=m_c^2$}
&\multicolumn{3}{l}{~~~~~~~~~~~~~$\mu^2=10$~GeV$^2$} \\

~datasets~
    & ~~~~$\langle x \rangle_v^\pi$ 
    & ~~~~$\langle x \rangle_s^\pi$ 
    & ~~~~$\langle x \rangle_g^\pi$

    & ~~~~$\langle x \rangle_v^\pi$ 
    & ~~~~$\langle x \rangle_s^\pi$ 
    & ~~~~$\langle x \rangle_g^\pi$          \\ \hline
~DY~  
    & ~~0.59(1)
    & ~~0.28(10)
    & ~~0.13(11)

    & ~~0.49(1)
    & ~~0.26(8)
    & ~~0.25(8)~                         \\

~DY+LN~
    & ~~0.53(2)
    & ~~0.14(4)
    & ~~0.34(6)    

    & ~~0.43(2)
    & ~~0.17(3)
    & ~~0.40(4)~                          \\    

~DY+LN+DY$\pT$~
    & \hspace*{0.22cm}{\bf 0.54(2)}
    & \hspace*{0.22cm}{\bf 0.16(3)}
    & \hspace*{0.22cm}{\bf 0.29(5)}

    & \hspace*{0.22cm}{\bf 0.44(1)}
    & \hspace*{0.22cm}{\bf 0.19(2)}
    & \hspace*{0.22cm}{\bf 0.37(3)}~    \\  \hline
\end{tabular}
\label{t.msr}
\end{table}

The addition of the $\pT$-dependent Drell-Yan data, for the fit labeled as ``full'' in Fig.~\ref{f.xpi}, does not affect the momentum fraction appreciably, resulting in a slight reshuffling of strength between the sea quark and gluon sectors, but within the $1\sigma$ uncertainties, with the valence fraction essentially unchanged.
The final momentum fractions for the full fit for the valence quark, sea quark and gluon contributions at the input scale $\mu^2=m_c^2$ are 0.54(2), 0.16(3) and 0.29(5), respectively.
Table~\ref{t.msr} summarizes the momentum fractions for all three scenarios at both the input scale and at the scale $\mu^2=10$~GeV$^2$.
Notably, when the PDFs are evolved to the higher scale, the momentum fraction carried by valence quarks decreases by $\sim 0.1$, whereas the gluon momentum fraction increases considerably, and a smaller increase occurs for sea quarks.
The uncertainties on the moments decrease as the moments are evolved to the higher scale.

\subsection{Flavor decomposition of observables}
\label{ssec.flavordecomp}

A deeper understanding of the impact on the pion PDFs from the various observables considered in our analysis can be obtained by examining the relative flavor contributions to the observables, especially from the less well constrained sea quark and gluon distributions.
In particular, we wish to study how the different flavors build up the LN structure function, which is sensitive to PDFs at small $x$, and the $\pT$-differential Drell-Yan cross section, in which the gluon PDF enters at the lowest order through the $qg$ channel at ${\cal O}(\alpha_s)$.
In \fref{channels} we show the valence quark, sea quark and gluon contributions to the $\pT$-integrated and $\pT$-differential Drell-Yan cross sections and to the LN structure function, relative to the total cross sections, at selected kinematics.

Firstly, for the $\pT$-integrated Drell-Yan cross section, $\diff^2\sigma^{\mbox{\tiny \rm DY}}/\diff \sqrt{\tau} \diff x_F$, shown in \fref{channels} versus $x_F$ at fixed $\sqrt{\tau}=0.288$, the most striking observation is the near dominance of the valence quark contribution to the total cross section across the entire $x_F$ range.
The sea quark contribution grows at small $x_F$, but does not exceed $\approx 15\%$ of the total, while the gluon component is almost negligible over all $x_F$.
Analyses that include only the $\pT$-integrated Drell-Yan data yield valence quark PDFs that are reasonably well determined, but leave the sea quark and gluon distributions almost totally unconstrained.

\begin{figure}[t]
\centering
\includegraphics[width=0.999\textwidth]{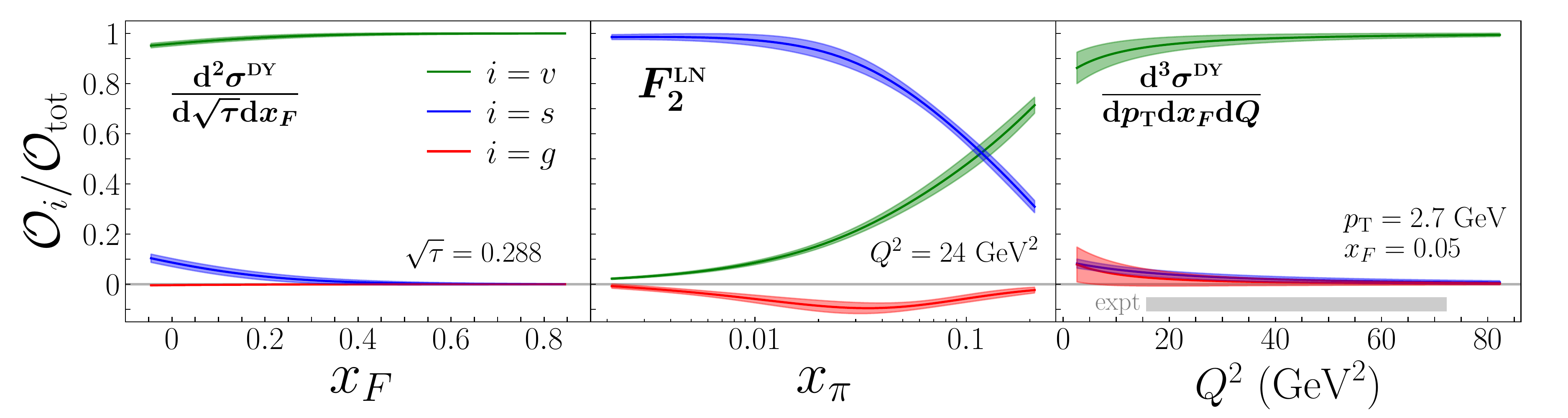}
\caption{Contributions from valence quarks ($i=v$, green bands), sea quarks ($i=s$, blue bands) and gluons ($i=g$, red bands) to observable ${\cal O}_i$ relative to the total, ${\cal O}_{\rm tot}$, for
    {\bf (left)}
Drell-Yan $\pT$-integrated cross section $\diff^2\sigma^{\mbox{\tiny \rm DY}}/\diff \sqrt{\tau} \diff x_F$ versus $x_F$ at $\sqrt{\tau}=0.288$,
    {\bf (middle)}
LN structure function $F_2^{\LN}$ versus $x_\pi$ at $Q^2=24$~GeV$^2$,
    {\bf (right)}
triply-differential $\pT$-dependent Drell-Yan cross section
$\diff^3\sigma^{\mbox{\tiny \rm DY}}/\diff\pT \diff x_F \diff Q$ versus $Q^2$ at $\pT=2.7$~GeV and $x_F=0.05$. The gray horizontal band represents the range of $Q^2$ for current and upcoming pion-induced DY measurements.}
\label{f.channels}
\end{figure}

As observed in \fref{PDFuncertainties}, the greatest impact on the sea quark and gluon distributions is from the HERA $F_2^{\rm LN}$ LN structure function data, shown in Fig.~\ref{f.channels} versus $x_\pi$ at a representative value of $Q^2=24$~GeV$^2$.
Interestingly, the dominant contribution to $F_2^{\rm LN}$ at low $x_\pi$ ($x_\pi \lesssim 0.01$) is from sea quarks, while the valence quark contributions become larger only at $x_\pi \gtrsim 0.1$.  
The gluon contribution is strongly suppressed across all $x_\pi$, largely because it appears only at NLO, and at intermediate $x_\pi$ is actually negative.
However, since with the Drell-Yan and LN data both the valence quark and sea quark PDFs become more strongly constrained, as \fref{PDFuncertainties} indicates, the uncertainties on the gluon PDF also decrease because of the momentum sum rule.
Thus the LN and Drell-Yan data play rather complementary roles in constraining valence and sea sectors of the pion.

More direct constraint on the gluon PDF is expected from the $\pT$-differential Drell-Yan cross section, which, because of the large $\pT$ of the exchanged virtual photon involved in the process, requires a hard gluon radiation (see Fig.~\ref{f.DYpTLO}).
In Fig.~\ref{f.channels}, we show the ratio of flavor contributions to the triply differential Drell-Yan cross section, $\diff^3\sigma^{\mbox{\tiny \rm DY}}/\diff\pT \diff x_F \diff Q$, versus $Q^2$ at fixed $x_F=0.05$ and at the minimal transverse momentum of the E615 data used in this analysis, namely $\pT=2.7$~GeV.
Note that triply differential cross section data for the pion-induced Drell-Yan process currently do not exist, but are shown here simply to illustrate the kinematics.

The general trend is qualitatively similar to the $\pT$-integrated Drell-Yan data, with the valence contributions dominating the cross section over all $Q^2$.
However, the gluon contribution, while comparatively small, is significantly larger here than for the $\pT$-integrated data, increasing to $\sim 10\%$ at the lowest $Q^2$ values. 
The horizontal band at the bottom of the figure represents the limits on current and future Drell-Yan experiments at $4 < Q < 8.5$~GeV, based on past experiments, such as E615 at Fermilab~\cite{E615}, and projections for the future COMPASS++/AMBER experiment at CERN~\cite{Denisov:2018unj}.
Accessing and interpreting lower-$Q^2$ data in terms of collinear factorization is problematic, however, since here meson resonances appear, which do not lend themselves to simple partonic interpretations.
Nevertheless, the $\pT$-dependent Drell-Yan data are the most challenging to describe, as \fref{DYpTdot} illustrates, and more precise data on the $\pT$ dependence, especially for the triply differential cross section, data would be valuable~\cite{Denisov:2018unj}.

\subsection{Scale dependence}
\label{ssec.scale}

Before concluding the discussion of the phenomenological results of our analysis, we review the problem of factorization scale setting for the $\pT$-dependent Drell-Yan data.
Inclusive processes, such as deep-inelastic scattering and ($\pT$-integrated) Drell-Yan lepton-pair production, are based on a single hard scale that characterizes the reaction, usually taken to be the invariant mass of the exchanged virtual photon.
In those cases, the scale dependence, introduced in QCD factorization to evaluate the short-distance cross section and the evolution of the PDFs, is chosen to be the hard scale in order to optimize the perturbative convergence.
In contrast, a second scale enters when considering cross sections differential in $\pT$, and the scale choice for the factorization at large transverse momentum can be chosen to be either set by $\pT$ or by $Q$.

Formally, the choice of scale should not affect the results appreciably, since the $\pT$-dependent cross sections at large $\pT$ are factorized under the condition that $\pT \sim Q$.
In practice, however, using a lowest order perturbative calculation for the large transverse momentum region does not guarantee that the resulting cross section will be insensitive to the choice of scale.
It is important, therefore, when performing an analysis with lowest order accuracy, to vary the scale around the two hard scales in the problem in order to estimate the sensitivity of the extracted PDFs to the scale choice.

We performed fits to the $\pT$-dependent Drell-Yan data by selecting several different scale choices, ranging from $\mu=\pT/2$ to $\mu=2\pT$ and including $\mu=Q$, as illustrated in \fref{scales}.
The scale dependence was found to correlate strongly with the normalization of the datasets, but the overall effect on the extracted PDFs was mild.
The best agreement with the $\pT$-dependent Drell-Yan data was found for the smallest value of the scale, $\mu=\pT/2$, for which $\chi^2_{\rm dat} = 0.94$ and the normalization parameter $n_e = 0.83$ for the ($Q, \pT$)-dependent and $n_e = 0.50$ for the ($x_F, \pT$)-dependent Drell-Yan data.
The $\chi^2$ deteriorates with increasing values of the scale, with the fit for $\mu=\pT$ yielding $\chi^2_{\rm dat} = 2.11$, while for $\mu=2\pT$ giving $\chi^2_{\rm dat} = 3.63$ for the $\pT$-dependent data.
When using the scale $\mu=Q$, the quality of the fit is also worse, with $\chi^2_{\rm dat} = 3.18$.
The PDFs with the scale $\mu=2\pT$ are almost identical to those for $\mu=\pT$, and are not shown in \fref{scales}.
The fitted normalization factors, $n_e$, for all the scale settings other than $\mu=\pT/2$ were found to be away from unity and closer to 0.5, which was the minimum value that we have allowed in our analysis.
The correlation between the scale dependence and the fitted normalization factor indicates potentially the need to include higher-order corrections that can restore the normalization factor to be closer to unity, but we leave such studies for future work.

\begin{figure}[t]
\centering
\includegraphics[width=\textwidth]{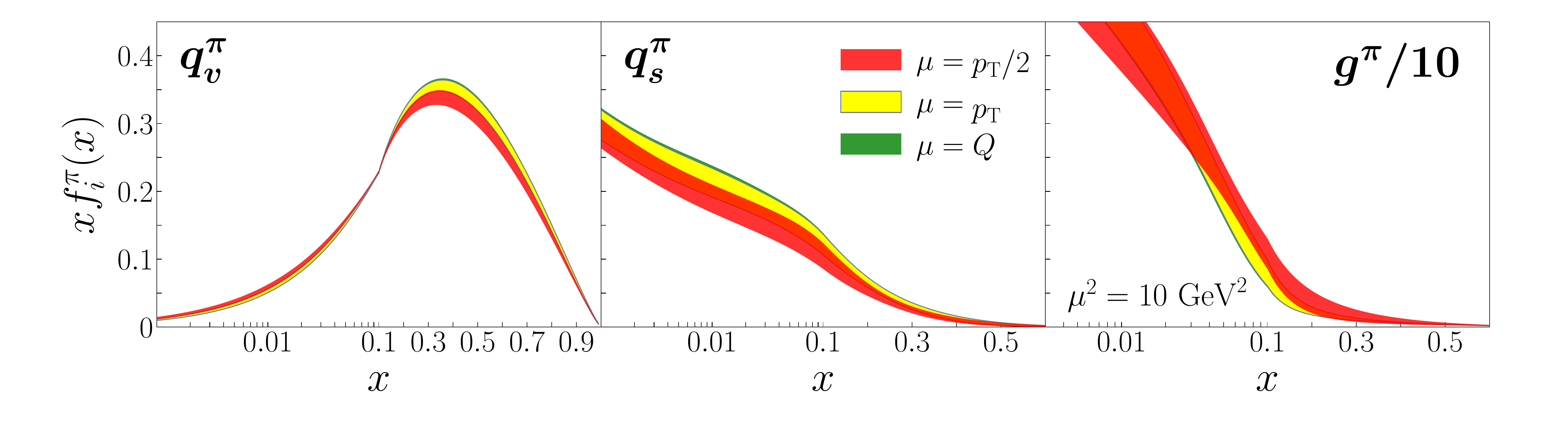}
\vspace*{-1cm}
\caption{Impact on the fitted pion valence quark $q_v^\pi$, sea quark $q_s^\pi$, and gluon $g^\pi$ (scaled by 1/10) PDFs by the choice of input scale $\mu$ for the $\pT$-dependent Drell-Yan data, for $\mu=\pT/2$ (red bands), $\mu=\pT$ (yellow bands) and $\mu=Q$ (green bands), at $\mu^2=10$~GeV$^2$. The bands are mostly overlapping with each other, indicating the PDFs are mostly insensitive to these choices of scale.}
\label{f.scales}
\end{figure}

\section{Contrasting pion and proton structure}
\label{s.PionVsProton}

As the lightest 3-valence quark and lightest quark-antiquark states, the proton and pion play a special role in nuclear physics, and understanding similarities as well as differences between their quark and gluon distributions can provide insights into fundamental aspects of hadron structure in QCD.
While the shapes of the valence $u$ and $d$ quark PDFs in the pion and proton are not expected to be the same because of the different valence contents and normalizations, a fascinating question is whether their sea quark content, and especially the gluon, are different.
Attempts at comparing the inclusive proton structure function and the pion structure function extracted from neutron electroproduction were made in the HERA analyses of LN cross sections~\cite{H1, ZEUS}, using data obtained under the same experimental conditions.
Here we revisit the proton versus pion structure function comparison in the context of a global QCD analysis, contrasting the results from the current pion analysis with the recent JAM19 proton PDFs obtained using  similar Monte Carlo methodology~\cite{JAM19}.

\subsection{Parton distributions}

\begin{figure}[t]
\centering
\includegraphics[width=0.8\textwidth]{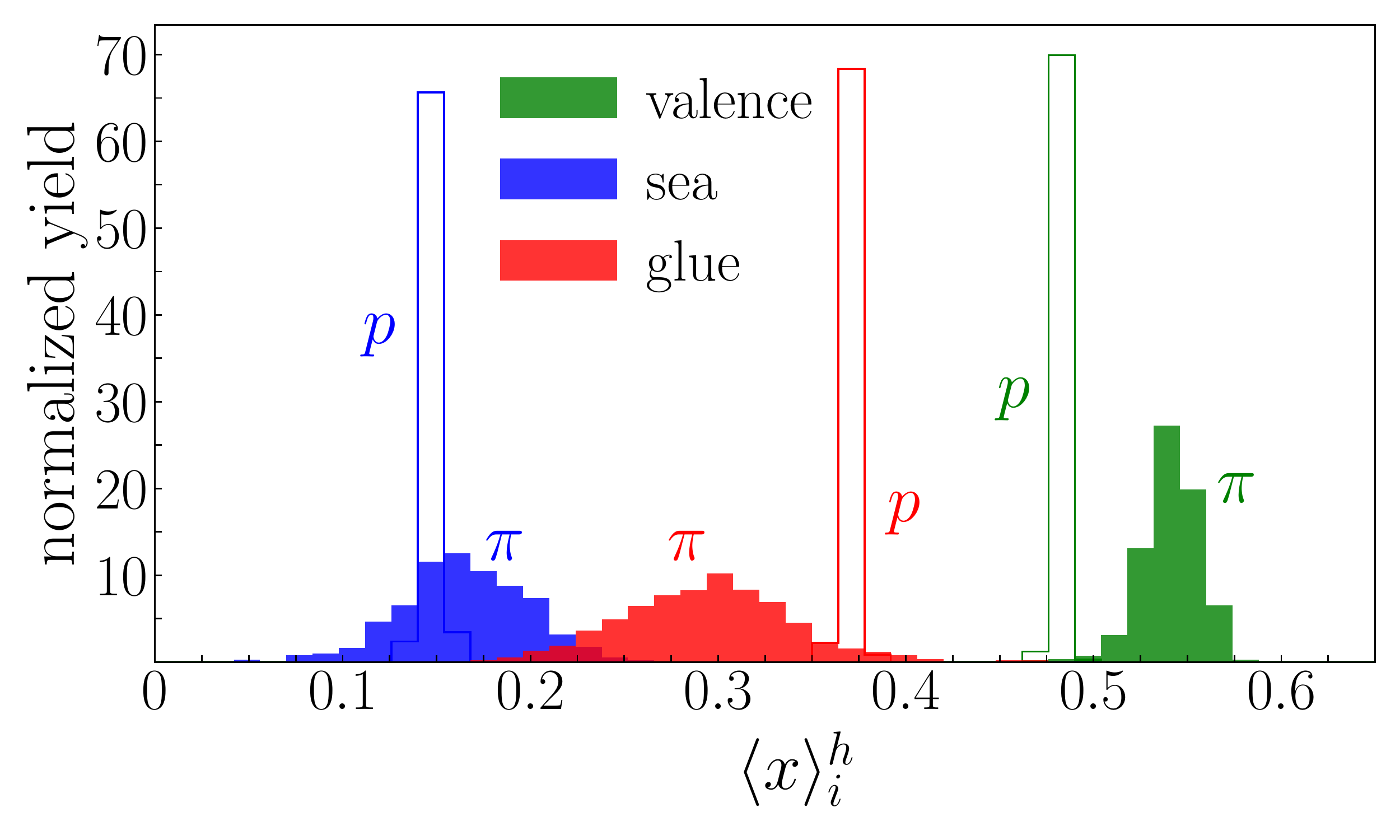}
\vspace*{-0.5cm}
\caption{Normalized yield of momentum fractions $\langle x \rangle_i^h$ of the proton ($h=p$, open) and pion ($h=\pi$, shaded), evaluated at the input scale, $\mu^2=m_c^2$, for valence quarks (green), sea quarks (blue), and gluons (red).
The proton PDFs used in calculating $\langle x \rangle_i^p$ are taken from Ref.~\cite{JAM19}.}
\label{f.pvspimsr}
\end{figure}

While the valence quark PDFs in the proton and pion are normalized differently, the momentum sum rule should hold for both hadrons, so that the momentum fractions carried by valence quarks, sea quarks and gluons can be compared directly.
In Fig.~\ref{f.pvspimsr}, the momentum fractions carried by each of these species in the proton and in the pion are shown in the form of histograms for the normalized yield, evaluated at the input scale, $\mu^2=m_c^2$.
The pion fractions correspond to the ``full'' results in Fig.~\ref{f.xpi}, while the proton fractions are computed from the JAM19 proton PDFs~\cite{JAM19}, with mean values and uncertainties given by
\begin{eqnarray}
\langle x \rangle^p_v &=& 0.481(3),\quad\quad
\langle x \rangle^p_s\ =\ 0.147(4),\quad\quad
\langle x \rangle^p_g\ =\ 0.372(3),\qquad
[\mu^2=m_c^2] \qquad
\label{eq.p_momfrac}
\end{eqnarray}
for the valence, sea and gluon, respectively at the input scale.
The most striking difference between the pion and proton results is the significantly narrower distributions for the latter, indicating much smaller uncertainties, about an order of magnitude, on the proton PDFs.
Interestingly, the valence quark momentum fraction in the proton is $\sim 10\%$ smaller than that in the pion (which is $\approx 3\sigma$, with the small proton uncertainties), while the central value of the gluon momentum fraction in the proton is $\sim 25\%$ larger, although compatible at the $\approx 1.5\sigma$ level with the much larger gluon pion uncertainties.
The total momentum carried by sea quarks is $\approx 15\%$ for both hadrons, with $\approx 7$ times smaller uncertainty for the proton.
For the case when the scale is evolved to $\mu^2=10$~GeV$^2$, the proton fractions are given by
\begin{eqnarray}
\langle x \rangle^p_v &=& 0.396(2),\quad\quad
\langle x \rangle^p_s\ =\ 0.181(3),\quad\quad
\langle x \rangle^p_g\ =\ 0.424(2),\qquad
[\mu^2=10~{\rm GeV}^2] \qquad
\label{eq.p_momfrac}
\end{eqnarray}
and as was the case in the pion, the valence quark momentum fraction decreases, while the gluon fraction increases more so than the sea quark with evolution.

\begin{figure}[t]
\centering
\includegraphics[width=0.95\textwidth]{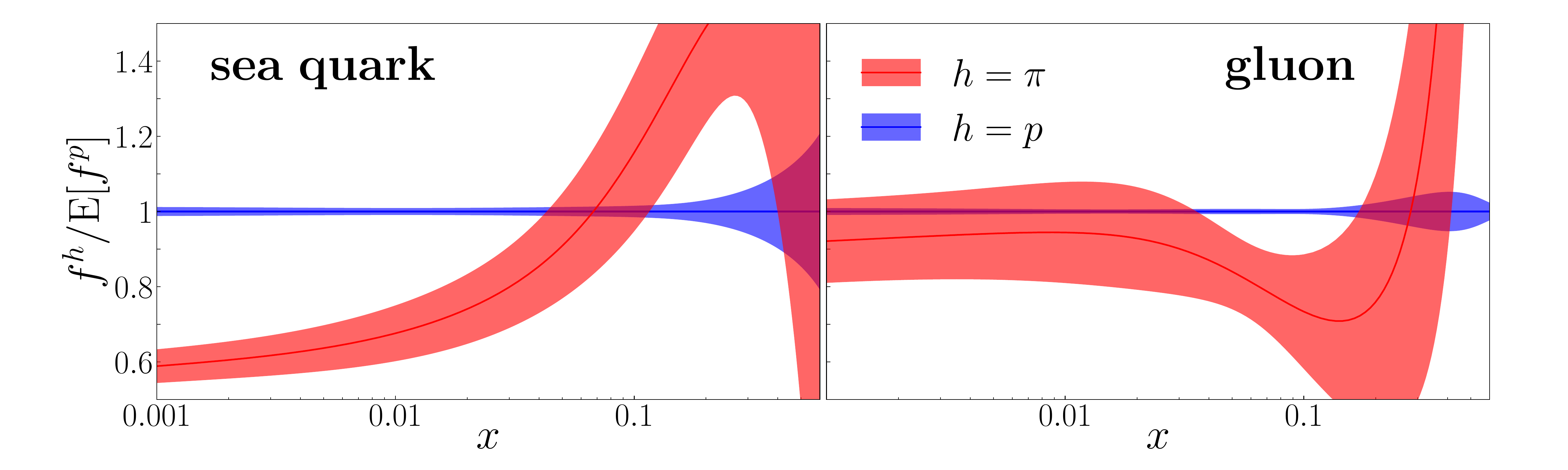}
\vspace*{-0.5cm}
\caption{Ratios of the sum of the sea quark {\bf (left)} and gluon {\bf (right)} PDFs in the pion (red bands) and proton (blue bands) to the mean value of the proton's distributions, evaluated using \eref{EV}, at a scale $\mu^2=10$~GeV$^2$. The bands represent 1$\sigma$ uncertainties on the PDFs.}
\label{f.pvspiglusea}
\end{figure}

The larger uncertainties on the $x$ dependence of the pion PDFs compared to those for the proton are illustrated in Fig.~\ref{f.pvspiglusea}, where we show the pion and proton total sea quark and gluon densities 1$\sigma$ uncertainty bands relative to the mean values of the {\it proton} PDFs, at a scale $\mu^2=10$~GeV$^2$.
Although the momentum fractions carried by sea quarks in the pion and proton are similar, as witnessed in Fig.~\ref{f.pvspimsr}, the shapes of the PDFs themselves are rather different.
In particular, the pion sea is significantly harder than the proton sea, with the latter $\sim 30\%-40\%$ larger for $x \lesssim 0.01$.
The pion and proton sea quark PDFs are comparable at $x \sim 0.1$, above which the pion's sea quark distributions become larger.
The larger sea quark PDFs at low $x$ in the proton has implications for the structure function comparison, as discussed below.

For the gluon distribution, the comparison in Fig.~\ref{f.pvspimsr} shows a greater degree of similarity between the pion and proton, with the small-$x$ behavior, $x \lesssim 0.04$, consistent within uncertainties.
At larger momentum fractions, $x \gtrsim 0.05$, on the other hand, the PDFs are noticeably different, though not to the extent of the sea quark distributions.
In particular, the pion gluon PDF dips below the proton gluon distribution, trending larger around $x \sim 0.2$.
A clear contrast between the two gluon distributions is the uncertainty associated with the PDFs across all $x$.
The proton's gluon PDF is much better constrained by data over all $x$ values than the pion's gluon PDFs, which reflects the considerably larger volume of high-energy scattering data, and calls for future pion data that could provide better constraints especially in the low-$x$ region.

\subsection{Structure functions}
\label{ssec.stfunc}

The neutron electroproduction experiments at HERA measured both the triply differential cross section 
    $\diff^3\sigma^{\LN}/\diff\xb \diff Q^2 \diff x_L$~\cite{H1}
and the ratio $r$ of the LN to inclusive cross sections~\cite{ZEUS}, as in Eq.~(\ref{e.ZEUSrxsec}).
The cross section is related to the LN structure function $F_2^{\LN}$ by the same kinematic factor in Eq.~(\ref{eq.LNxsec}) as for the inclusive cross section, 
    $\diff^2\sigma/\diff\xb \diff Q^2$.
Both the H1 and ZEUS experiments reported that the ratio $r$ was approximately a function of $x_L$ only, which suggested that $F_2^{\LN}(\xb,Q^2,x_L) \propto F_2^p(x_B,Q^2)$ at a given $x_L$, and the proportionality was shown to hold for intermediate to high values of $x_L$.

In the factorized approximation (\ref{eq.F2LN3}) in terms of pion exchange, the $F_2^{\LN}$ structure function is expressed in terms of the pion-nucleon splitting function $f_{\pi N}(x_L)$ and the pion structure function $F_2^\pi(x_\pi,Q^2)$, with $x_\pi=\xb/\bar{x}_L$.
For the ratio $r$ to be a function of $x_L$ only, the $x_B$ (and $Q^2$) dependence in the numerator and denominator need to cancel, and since the splitting function depends only on $x_L$, the $F_2^\pi$ and $F_2^p$ structure functions would be proportional at a given $x_L$, with a proportionality factor $\alpha$,
    $F_2^\pi(x_\pi,Q^2) \approx \alpha F_2^p(\xb,Q^2)$.
Using the results from our analysis, we are able to confront directly whether the equality holds for a scaled proton structure function by computing $F_2^p(x_B,Q^2)$ and $F_2^\pi(x_\pi,Q^2)$ explicitly.

\begin{figure}
\centering
\includegraphics[width=0.65\textwidth]{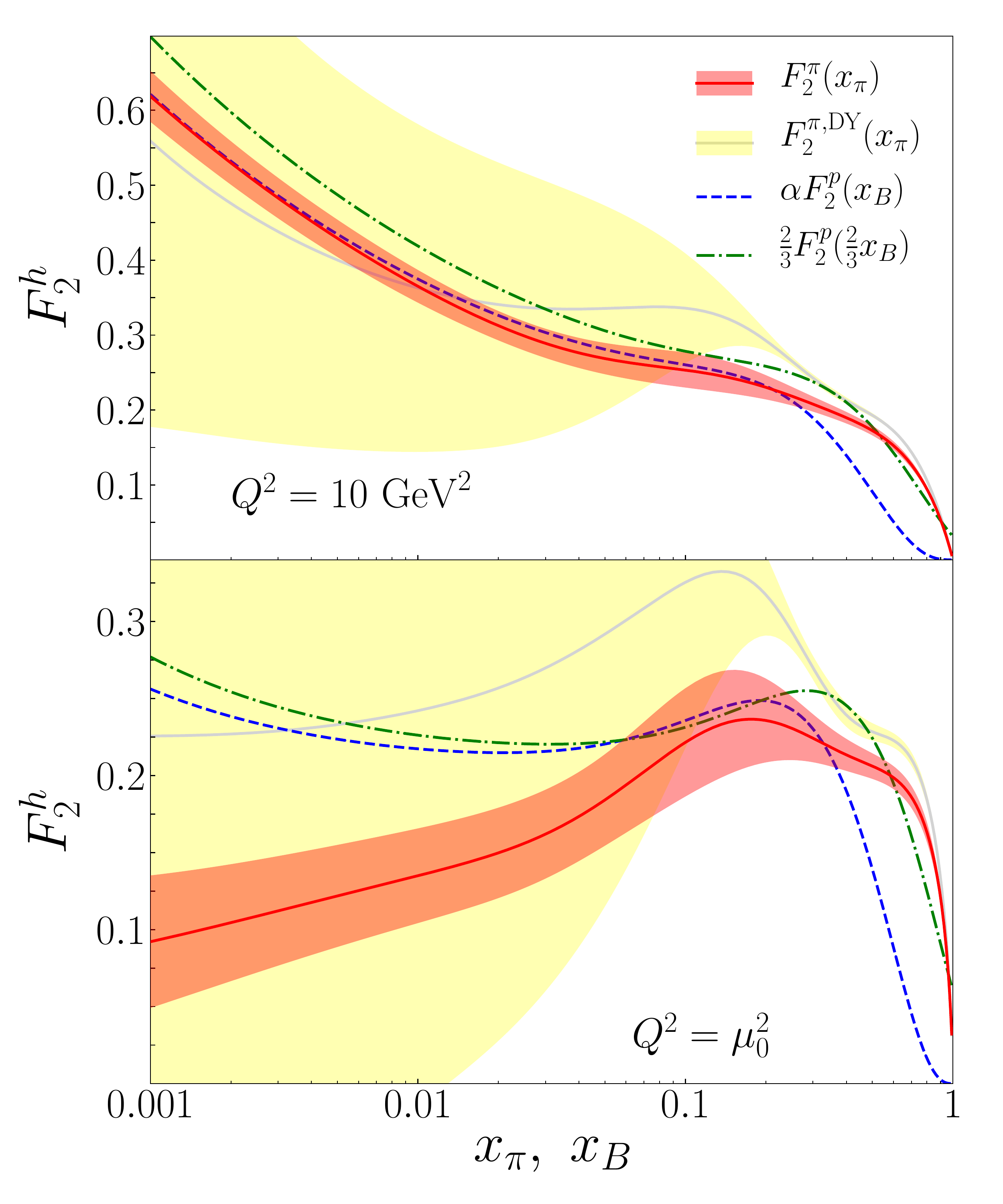}
\vspace*{-0.5cm}
\caption{Comparison of the pion structure function from the full analysis, $F_2^\pi$ (red lines and 1$\sigma$ uncertainty bands) and with Drell-Yan data only, $F_2^{\pi, \rm DY}$ (gray lines and 1$\sigma$ yellow uncertainty bands) with models of $F_2^\pi$ based on the proton structure function, evaluated at $Q^2=10$~GeV$^2$ {\bf (top)} and at the input scale $Q^2=\mu_0^2=m_c^2$ {\bf (bottom)}. The models include $F_2^p$ scaled by a factor $\alpha=0.65$ (blue dashed lines) and scaled as $\frac23 F_2^p(\frac23\xb)$~\cite{NSZ} (green dot-dashed lines).}
\label{f.F2comp}
\end{figure}

An alternative model that gives reasonable agreement with the LN data is the color-dipole model, based on the BFKL-Regge expansion, introduced by Nikolaev {\it et al.} (NSZ)~\cite{NSZ}, which approximates the pion structure function by the scaled proton structure function at a shifted value of $\xb$,
    $F_2^\pi(x_\pi,Q^2) \approx \frac23 F_2^p(\frac23 \xb,Q^2)$.
Here, the 2/3 normalization factor comes from the ratio of the pion to proton eigenfunctions of the color-dipole BFKL equation, whereas the $x_\pi$ dependence comes from an inverse proportionality of the effective dipole-dipole scattering energy, a quantity which is $3/2$ times larger in the pion than in the proton because of the number of available valence quarks.

In \fref{F2comp} we compare the $x_\pi$ dependence of the pion structure function from our global QCD analysis with the two models in which $F_2^\pi$ is computed from the proton $F_2^p$ structure function, both at the input scale $Q^2=\mu_0^2$ and evolved to $Q^2=10$~GeV$^2$.
For the simple scaling model we fix the proportionality constant at the value $\alpha=0.65$ fitted to give the best agreement at $Q^2=10$~GeV$^2$, which happens to be similar to the 2/3 factor of the NSZ model.
We also show the pion structure function from the analysis including only Drell-Yan $\pT$-integrated data, $F_2^{\pi,{\rm DY}}$.
As the uncertainties on $F_2^{\pi,{\rm DY}}$ are very large at low $x_\pi$, the inclusion of the LN data in our analysis is vital to obtain better constraint on the pion's structure function.
We also note that since no data currently exist for pion structure functions at $Q^2=m_c^2$, the $F_2^\pi$ in Fig.~\ref{f.F2comp} at the input scale involves an element of extrapolation.

Interestingly, at $Q^2=10$~GeV$^2$ the scaled proton structure function $\alpha F_2^p(x_B,Q^2)$ follows closely the pion structure function at small values of $x_\pi \lesssim 0.2$ and lies within its uncertainties.
At large $x_\pi$ and $\xb$, on the other hand, the pion $F_2^\pi$ has a relatively slow falloff as $x_\pi \to 1$, while the proton $F_2^p$ is much softer and falls faster as $\xb \to 1$.
This effect simply reflects the different behaviors of the valence quark PDFs in the pion and proton.
In contrast, for the NSZ model the large-$\xb$ behavior of the rescaled $F_2^p$ is more consistent with the pion structure function, due to the 2/3 rescaling factor in the argument.
However, at lower $\xb$ the agreement with the pion $F_2^\pi$ is not as good in this model.
These observations suggest that a hybrid model, in which the low-$\xb$ behavior ($\xb \lesssim 0.2$) is given by $\alpha F_2^p(\xb,Q^2)$ and the high-$\xb$ behavior ($\xb \gtrsim 0.2$) is given by $\frac23 F_2^p(\frac23 \xb,Q^2)$, may be a better representation of~$F_2^\pi$.

Note that whatever agreement exists between the proton and pion structure functions at $Q^2=10$~GeV$^2$ does not translate to lower $Q^2$ values.
At the input scale, $Q^2=\mu_0^2$, for example, the shapes of $F_2^\pi$ and the two models based on $F_2^p$ differ substantially, with opposite slopes.
In particular, at low values of $\xb \lesssim 0.1$ the latter are significantly larger than the $F_2^\pi$ from the global analysis, with neither model able to  accurately capture the low-$x_\pi$ behavior of $F_2^\pi$.

The agreement between $F_2^\pi$ and $F_2^p$ for low $x_\pi$ and $\xb$ at $Q^2=10$~GeV$^2$, and the disagreement at $Q^2=\mu_0^2$, can be understood in terms of the flavor decomposition of the structure functions into valence, sea, and gluon components and the different roles played by each.
Firstly, since the valence quark PDFs in the pion and proton have rather different $x$ dependence, the structure functions at large $x_\pi$ or $\xb$ are expected to differ.
At low $x$ values, the sea quark PDFs in the pion and proton are also quite different, with a nontrivial $x$ dependence, as illustrated in Fig.~\ref{f.pvspiglusea}.
This is directly responsible for the difference between $F_2^\pi$ and $F_2^p$ at low $x_\pi$ and $\xb$ at the input scale.
As $Q^2$ increases, the role of evolution becomes more important, and the contributions to the structure functions from gluons become more prominent, even though formally these enter at higher order in $\alpha_s$.
Since gluon PDFs in the pion and proton are fairly similar at low values of $x$, as observed in Fig.~\ref{f.pvspiglusea}, their contributions to the structure functions at low $x_\pi$ and $\xb$ must also be similar, which is reflected in the better agreement between the pion and proton $F_2$ functions at higher $Q^2$ values.
Overall, however, our results would suggest that the structure functions at low $x_\pi$ and $\xb$ are strongly dependent on the nonperturbative sea quark distribution and on the $Q^2$ evolution of the PDFs, and it is difficult to draw general, scale-independent conclusions about their characteristics.

\section{Outlook}
\label{s.Outlook}

One of the most important achievements of the present work has been the demonstration that it is indeed possible to describe, within the same $\mathcal{O}(\alpha_s)$ collinear factorization framework, the $\pT$-dependent spectrum of Drell-Yan lepton pairs in pion--nucleon collisions, along with the more traditional $\pT$-integrated Drell-Yan data and LN electroproduction cross sections.
To our knowledge, this is the first demonstration of its kind for the pion, and paves the way to the quantitative study of the three-dimensional parton structure of the pion.
The inclusion of the $\pT$-dependent data contributes to the reduction of uncertainties on the gluon distribution at large $x$, although the impact of the existing data is relatively modest compared to the $\pT$-integrated Drell-Yan and LN data, and calls for future, high-precision measurements of large-$\pT$ pion-induced lepton-pair production cross sections.

Future improvements to this work will implement threshold resummation in the perturbative calculations of $\pT$-integrated Drell-Yan cross sections~\cite{PionResum}, which can produce non-negligible enhancement of the cross section from the emission of soft gluons near threshold.
Previous work~\cite{Aicher} indicated potentially significant impact on the large-$x$ behavior of the valence pion PDF, and it will be important to explore the systematic uncertainties of the calculation~\cite{Catani96, Bonvini11, Westmark17}.
Beyond this, future theoretical improvements in the analysis of $\pT$-dependent data should incorporate $\pT$ smearing as well as higher-order gluon emissions, both of which affect the energy and kinematics of incoming partons.

On the experimental side, the planned program of pion- and kaon-induced Drell-Yan measurements at the future COMPASS++/AMBER~\cite{Denisov:2018unj} facility at CERN should provide precision data to better determine the gluon and sea quark PDFs in the pion, and allow first glimpses of the partonic structure of kaons.
Additional $\pT$-dependent Drell-Yan data at high $\pT$ may help to constrain the valence quark and gluon distributions at large values of~$x$.
Other planned experiments, such as the Tagged Deep-Inelastic Scattering (TDIS) experiment at Jefferson Lab, will provide complementary information to the HERA LN data by measuring a leading proton produced in DIS from a neutron in a deuteron.
Utilizing the Sullivan process, this reaction will probe the structure of the exchanged pion in a kinematic region between the (low-$x$) HERA and (high-$x$) Drell-Yan domains.
Experiments at the future Electron-Ion Collider (EIC) may also provide better statistics for LN measurements at low~$x_\pi$ values in tagged electron-proton collisions.

\section*{ACKNOWLEDGEMENTS}

We thank C.-R.~Ji and A.~Radyushkin for helpful discussions, and I.~Novikov and S.~Glazov for assistance with the xFitter results.
This work was supported by the US Department of Energy contract DE-AC05-06OR23177, under which Jefferson Science Associates, LLC operates Jefferson Lab, by the US Department of Energy contract DE-FG02-03ER41260, and by the US Department of Energy, Office of Science, Office of Workforce Development for Teachers and Scientists, Office of Science Graduate Student Research (SCGSR) program. The SCGSR program is administered by the Oak Ridge Institute for Science and Education for the DOE under contract number DE‐SC0014664.


\end{document}